\documentclass[aps,prb,amsmath,twocolumn,amssymb,floatfix,nosuperscriptaddress,nofootinbib]{revtex4-2}
\usepackage{physics} 
\usepackage{orcidlink}
\usepackage{tikz} 
\usepackage{lineno}
\usepackage{graphicx} 
\usepackage{bm}

\newcommand{\vf}{v_{\textrm{F}}}
\newcommand{\vh}{v_{\textrm{H}}}
\newcommand{\vnh}{v_{\textrm{NH}}}
\newcommand{\Ts}{T_{\textrm{S}}}
\newcommand{\Tas}{T_{\textrm{AS}}}
\newcommand{\vB}{v_{\textrm{B}}}
\newcommand{\bfk}{\mathbf{k}} 
\newcommand{\bfq}{\mathbf{q}} 

\begin{document}

\title{Yukawa-Lorentz Symmetry of Tilted Non-Hermitian Dirac Semimetals at Quantum Criticality}

\author{Sergio Pino-Alarcón}
\affiliation{Departamento de Física, Universidad Técnica Federico Santa María, Casilla 110, Valparaíso, Chile}

\author{Vladimir Juri\v{c}i\'c}\thanks{Corresponding author: vladimir.juricic@usm.cl}
\affiliation{Departamento de Física, Universidad Técnica Federico Santa María, Casilla 110, Valparaíso, Chile}

\begin{abstract}
Dirac materials, hosting linearly dispersing quasiparticles at low energies, exhibit an emergent Lorentz symmetry close to a quantum critical point (QCP) separating a semimetallic state from a strongly-coupled gapped insulator or superconductor. This feature appears quite robust even in open Dirac systems coupled to an environment, featuring non-Hermitian (NH) Dirac fermions: close to a strongly coupled QCP, a Yukawa-Lorentz symmetry emerges with a unique terminal velocity for both the fermion and the bosonic order parameter fluctuations. The system can either retain non-Hermiticity or completely decouple from the environment, recovering Hermiticity as an emergent phenomenon. We show that such Yukawa-Lorentz symmetry emerges at quantum criticality even when the NH Dirac Hamiltonian includes a tilt term at the lattice scale. As demonstrated by a leading-order $\epsilon=3-d$ expansion near the $d=3$ upper critical dimension, a tilt term becomes irrelevant near the QCP separating the NH Dirac semimetal and a gapped (insulating or superconducting) phase. This behavior extends to linear-in-momentum non-tilt perturbations, introducing velocity anisotropy for the Dirac quasiparticles, which also becomes irrelevant at the QCP. These predictions can be numerically tested in quantum Monte Carlo lattice simulations of NH Hubbard-like models hosting low-energy NH tilted Dirac fermions.
\end{abstract}

\maketitle

\section{Introduction}
\label{sec:intro}

Dirac materials, hosting linearly dispersing quasiparticles at low energies, exhibit an emergent relativistic-like Lorentz symmetry due to the  quasiparticles' linear energy-momentum dispersion, with the Fermi velocity, typically a few hundred times smaller than the velocity of light,   playing the role of the velocity of light~\cite{Neto-2009,Balatsky2014,armitage2018}.  This symmetry is surprisingly robust, persisting even with  electron-electron interactions, which break it at the lattice (UV) scale. For example, such a pseudorelativistic Lorentz symmetry emerges in the deep infrared regime for two-dimensional and three-dimensional Dirac fermions interacting via the long-range Coulomb interaction, featuring a common terminal velocity for both the Dirac fermions and the photons~\cite{Gonzalez1994, Isobe2012}.  Additionally, for Yukawa (short-range) interacting Dirac fermions, the quantum-critical point (QCP) that separates the semimetallic phase from a strongly-coupled gapped (insulating or superconducting) phase  also exhibits this symmetry~\cite{SSLee2007,HJV-PRB2009,Roy2016,Roy2018,Roy2020}.

Recent studies strongly suggest that this scenario also  extends to the non-Hermitian (NH) Dirac systems~\cite{jurivcic2024yukawa,Murshed2024,asrap2024yukawa}, pertaining to Dirac fermions coupled to an environment, described by a minimal Dirac operator capturing non-Hermiticity, as given by Eq.~\eqref{eq:NH-Dirac}. Such an NH Dirac operator, with its anti-Hermitian part  linear in momentum,  is  constructed by invoking a Dirac mass term in the Hermitian Dirac Hamiltonian,  represented by a Hermitian matrix $M$ that anticommutes with the Hermitian Dirac Hamiltonian~\cite{jurivcic2024yukawa}. It features a purely real or imaginary spectrum, with a linearly vanishing density of states (DOS) for a weak non-Hermiticity, as discussed in Sec.~\ref{subsec:DoS}, implying  the  stability of NH Dirac semimetal for weak short-range Hubbard-like electron-electron interactions. However, when such a short-range interaction, effectively described by the Yukawa coupling, is strong enough, the system undergoes a quantum-phase transition with the effective Fermi velocity of the  NH Dirac fermions and the bosonic order parameter (OP) excitations  reaching  a common value at a QCP. The fate of the coupling  with the environment, however, depends on the canonical commutation relations between the OP and the Dirac mass matrix $M$~\cite{jurivcic2024yukawa}: when the OP commutes with the matrix $M$, the latter therefore representing the commuting class mass (CCM) OP, the system retains non-Hermiticity, while featuring  a common velocity for all participating degrees of freedom, thus yielding  an emergent NH Yukawa-Lorentz symmetry. On the other hand,  in the case of an anticommuting mass (ACM) OP, the system completely decouples from the environment, thereby  restoring the usual Lorentz symmetry, with concomitant emergence of the  Hermiticity. Such behavior of NH (Hermitian) Dirac fermions suggests  ubiquity of the emergent Yukawa-Lorentz (Lorentz) symmetry at QCPs, despite being absent at the UV scale, and irrespective of the symmetry-breaking perturbations in the bare (lattice) Hamiltonian. This is also further supported by the emergence  of this symmetry for birefringent NH Dirac fermions, featuring different velocities, therefore breaking Lorentz symmetry at the bare level~\cite{Murshed2024}. 

From the point of view of the Dirac Hamiltonian, adding an operator linear in a single  component of the momentum  represents its minimal extension  that breaks the rotational, and therefore also Lorentz symmetry, which is, at the same time, marginal close to the semimetallic ground state. A particular  minimal non-Lorentz symmetric extension of the Dirac Hamiltonian is tilt of the Dirac Hamiltonian~\cite{Soluyanov2015}, represented by a matrix that commutes with it, which was shown to be irrelevant at the QCP in the Hermitian case~\cite{reiser2024tilted}. Such a tilt term appears in many instances when considering the Hermitian Dirac fermions, e.g.,  as a platform for simulating  curved spacetime in condensed matter systems~\cite{Volovik2016,Nissinen2017,Jafari-PRB2019,Ojanen-PRR2019,Jafari-PRR2020,Schmidt-SciPost2021,Volovik2021,Meng-PRR2022,vanWezel-PRR2022,vanderBrink-PRB2023}, it can exhibit nontrivial quantum transport~\cite{Bergholtz-PRB2015,Carbotte-PRB2016,Stoof-PRB2017,Goerbig-PRB2019,Rodriguez-Lopez2020,Maytorena-PRB2021,Hao-Ran-PRB2022,Portnoi-PRB2022,Goerbig-PRB2023}, while  the effects of the long- and short-range components of the Coulomb interaction~\cite{Fritz-PRB2019,Rostami-PRR2020,RudneiRamos-PRB2021,RudneiRamos-PRB2023,liu2024low}, various disorders~\cite{Bergholtz-PRB2017,Fritz-PRB2017,GZLiu-PRB2018,YWLee-PRB2019}, and the magnetic field~\cite{Goerbig-PRB2008,Carbotte-PRB-2018,Nag_2021}  have also been explored in this context. The tilt term may also exhibit nontrivial transport effects for NH Dirac models~\cite{qin2024nonlinear}. 
However, the role of this Lorentz-symmetry breaking perturbation and the ultimate fate of the tilt for quantum-critical NH Dirac fermions, in relation to the emergence of Yukawa-Lorentz symmetry, remain unexplored, which is the focus of this article.

To this end, we consider the simplest scenario: a Hermitian tilt applied to an NH Dirac fermion. Such a minimal model describing tilted NH Dirac fermions may describe an effective dynamics of the tilted Dirac materials coupled to an external environment, such as a (fermionic or bosonic) quasiparticles' bath, with a direct connection between a microscopic and effective model to be established  in future. Our main result is the emergence of the Yukawa-Lorentz symmetry in the vicinity of the strongly coupled QCP separating the NH Dirac semimetallic phase from an insulating CCM phase since the tilt term becomes \emph{irrelevant} in the vicinity of this QCP (Eqs.~\eqref{eq:beta-function-vh}-\eqref{eq:beta-function-vb} and  Fig.~\ref{fig: assym}), while remaining coupled to the environment.  To show it, we here perform a field-theoretical analysis of the strongly interacting NH tilted Dirac fermions within a leading-order renormalization group (RG) $\epsilon=3-d$-expansion, close to $d=3$ upper-critical dimension of the Gross-Neveu-Yukawa field theory, in which the Yukawa and the quartic OP couplings are marginal. For weak short-range electron interactions, the NH Dirac semimetal is stable, due to the linearly vanishing DOS close to the zero energy, analyzed in Sec.~\ref{subsec:DoS}, see also Fig.~\ref{fig: dos}. Furthermore, our calculation of the mean-field susceptibilities shows that the competition between the ACM and CCM OPs, in spite of the ACM OPs being favored at the mean-field level (Figs.~\ref{susc as} and~\ref{susc s}), depends on the system's parameters at the corresponding QCP.  This further motivates the analysis of the quantum-critical behavior using the RG formalism, which includes the effects of the fluctuations in the quantum-critical region, ultimately uncovering the Yukawa-Lorentz symmetry for the quantum-critical tilted NH Dirac fermions. Furthermore, the fate of the non-Hermiticity depends on the algebra of the NH Hamiltonian and the OP, more precisely, whether the OP belongs to the ACM or CCM class, as shown in Fig.~\ref{fig: assym}.  In the former case, the system completely decouples from the environment, therefore restoring the hermiticity, while for the latter, the system remains coupled to the environment while displaying an emergent Yukawa-Lorentz symmetry. For completeness, we also consider a term linear in momentum that neither commutes nor anticommutes with the NH Hamiltonian,  which introduces a velocity anisotropy, and as we demonstrate, it also becomes irrelevant close to the QCP and eventually vanishes (Fig.~\ref{fig: symm}), while the system decouples from the environment, restoring Hermiticity and Lorentz symmetry. 

Altogether, our results support a scenario of  the Yukawa-Lorentz symmetry being a universal feature of quantum-critical NH Dirac fermions. This prediction should be tested in the numerical simulations, such as quantum Monte Carlo, for instance, in the NH Hubbard model with the asymmetric  hoppings together with the tilt term implemented on the honeycomb lattice, analogously to the Hubbard model for the tilt-free NH Dirac fermions~\cite{Xiang-PRL2024}. 
\subsection{Organization}
The rest of the paper is organized as follows. In Sec.~\ref{sec:minimal model}, we introduce the model of the NH Dirac fermions with an extra term linear in momentum, which includes the tilt and the velocity anisotropy, and we calculate the DOS (Sec.~\ref{subsec:DoS}). Sec.~\ref{sec:susceptibilities} is dedicated to the analysis of the mean-field susceptibilities in $d=3$, while in Sec.~\ref{sec:NH-critical} we discuss the features of the quantum-critical Gross-Neveu-Yukawa theory for the tilted Dirac fermions. In particular, we calculate the RG flows of the fermionic and bosonic velocities and the tilt parameter, establishing the emergence of the Yukawa-Lorentz (Lorentz) symmetry at the QCP controlling the transition into a CCM (ACM) order. The conclusions are presented in Sec.~\ref{sec:conclusions}, while the various technical details are relegated to the appendices.

\section{Minimal noninteracting tilted non-Hermitian Dirac Hamiltonian}
\label{sec:minimal model}
We start by considering  a minimal effective low-energy Hamiltonian-like NH Dirac operator $H_{\textrm{NH}}$ in two spatial dimensions ($d=2$) which is constructed  from the corresponding Hermitian Dirac operator by invoking the mass terms, such that the resulting anti-Hermitian piece of the total (Lorentz-symmetric) NH Dirac operator is linear in  momentum~\cite{jurivcic2024yukawa}. The Hermitian Dirac Hamiltonian reads as
$\mathcal{H}_D=\sum_{\vb{k}}\Psi_{\vb{k}}^\dag H_{\textrm{D}}\Psi_{\vb{k}}$, where
\begin{equation}
\label{eq:HermitianDirac}
H_{\textrm{D}}=\vh\qty(\Gamma_x k_x + \Gamma_y k_y)=\vh h_0,
\end{equation}
with $\vh$ as the Hermitian velocity of the Dirac quasiparticles,  the Hermitian matrices $\Gamma_x=\sigma_1 \otimes \tau_3$ and $\Gamma_y=\sigma_2 \otimes \tau_0$, chosen here explicitly in this representation for convenience,  are mutually anticommuting matrices satisfying the  Clifford algebra  $\{\Gamma_i,\Gamma_j\}=2 \delta_{ij}$, $i,j=1,2$, $\vb{k}$ is the momentum, and the form of the Dirac spinor $\Psi_{\vb{k}}$ depends on the microscopic details of the system, which is in the following   assumed to be four-dimensional, as for instance for the spinless electrons in graphene, with two sublattice  and two valley degrees of freedom. The Pauli matrices $(\sigma_0, {\bm \sigma})$ and $(\tau_0, {\bm \tau})$ act in the sublattice  and valley space, respectively.

To construct the NH counterpart of the Dirac operator in Eq.~\eqref{eq:HermitianDirac}, we invoke a Hermitian  matrix $M$, corresponding to a mass term for the Hermitian Dirac operator, and therefore 
$\{M,H_{\textrm{D}}\}=0$. As such, for any mass matrix, the operator  $MH_{\textrm{D}}$ is anti-Hermitian, $(M H_{\textrm{D}})^\dagger=-M H_{\textrm{D}}$. Using this property, we define he minimal NH Dirac operator as~\cite{jurivcic2024yukawa}
\begin{equation}
\label{eq:NH-Dirac}
     H^0_{\textrm{NH}}=(\vh+\vnh M)h_0=\vh(1+\beta M)h_0\,
\end{equation}
where  $\vnh$ is the NH velocity parameter, and  $\beta=\vnh/\vh$ (always real).  
Then, the spectrum of the NH operator is $E_{\textrm{NH}}=\pm\vf \abs{\vb{k}}$, and is purely real (imaginary) when $|\beta|<1$ ($|\beta|>1$), where  $\vf=\vh\sqrt{1-\beta^2}$  is the effective Fermi velocity of the NH Dirac fermions. For concreteness, we here choose the mass matrix $M=\sigma_1\otimes\tau_1$, and the velocity parameter $|\beta|<1$, with the NH Dirac operator thus featuring a purely real spectrum. Our results are, however, independent of the matrix representation.

As the last step, to obtain the minimal tilted NH Dirac Hamiltonian, we also add a Hermitian term which is linear in one of the components of the momentum~\cite{Soluyanov2015}, taken to be $k_x$ for concreteness,
\begin{equation}
\label{eq:tilt-NH-Dirac}
H_{\textrm{t}}^0=\alpha \vh T k_x,
\end{equation}
with the matrix $T$ such that $[T,\Gamma_i]=0$ ($i=x,y$) and $[T,\beta]=0$, corresponding to the tilted Dirac cones. Depending on the   spatial symmetries, particularly, the valley exchange symmetry, represented by $(k_x,k_y)\to(-k_x,k_y)$ and $S_V=\sigma_0\otimes\tau_1$, we can consider an antisymmetric and symmetric tilt with $\Tas=\sigma_0\otimes \tau_0$ and $\Ts=\sigma_2\otimes \tau_2$, respectively,  which is  odd and even under this symmetry, respectively.  The Hamiltonian of the minimal model for the NH tilted Dirac semimetal then reads,  
\begin{equation}
\label{eq:tilt-Hamiltonian}   H_{\textrm{NH}}^{\textrm{tilt}}(\vb{k})=(\vh+\vnh M)h_0+\alpha \vh T k_x.
\end{equation}
Notice that this Hamiltonian preserves and breaks  sublattice exchange symmetry, $(k_x,k_y)\to(k_x,-k_y)$ and $S_{SL}=\sigma_1\otimes\tau_0$ for any $\beta\neq0$, in the case of the antisymmetric and  symmetric tilt term, respectively. 

For completeness, we also consider the NH Dirac Hamiltonian that includes  the term $H_{\textrm{t}}^0$ in Eq.~\eqref{eq:tilt-NH-Dirac} with the matrix $T=T_{{\rm VA}}=\sigma_1\otimes\tau_0$, which is odd and even under the valley and sublattice exchange, respectivrly, and therefore transforms the same as the antisymmetric tilt under these symmetries. However, this term neither commutes nor anticommutes with the Hermitian Dirac Hamiltonian [Eq.~\eqref{eq:HermitianDirac}], $[\Gamma_x,T_{{\rm VA}}]=0,\, \{\Gamma_y,T_{{\rm VA}}\}=0 $, and thus  introduces a velocity anisotropy in  the NH Dirac electronic bands, see Eq.~\eqref{eq:spectrum-velocity-anisotropy}.  

In the case of antisymmetric  tilt, the spectrum of the NH tilted Hamiltonian  operator  features two doubly degenerate bands 
\begin{equation}
\label{eq:spectrum-asymmetric}
E_{\textrm{NH}}^{\textrm{AS}}=\pm \vf\abs{\vb{k}} + \alpha \vh k_x, 
\end{equation}
and therefore the tilt acts as an effective (valley-independent) momentum-dependent chemical potential for the NH Dirac fermions. On the other hand, for the symmetric  tilt, 
the spectrum reads as
\begin{equation}
E_{\textrm{NH}}^{\textrm{S}}=\pm \vf\abs{\vb{k}} \pm \alpha \vh k_x,
\end{equation}
and acts as an effective  momentum-dependent chiral chemical potential,  with an opposite sign at the two valleys. 

\begin{figure*}[t!]
        \centering
        \includegraphics[scale=0.39]{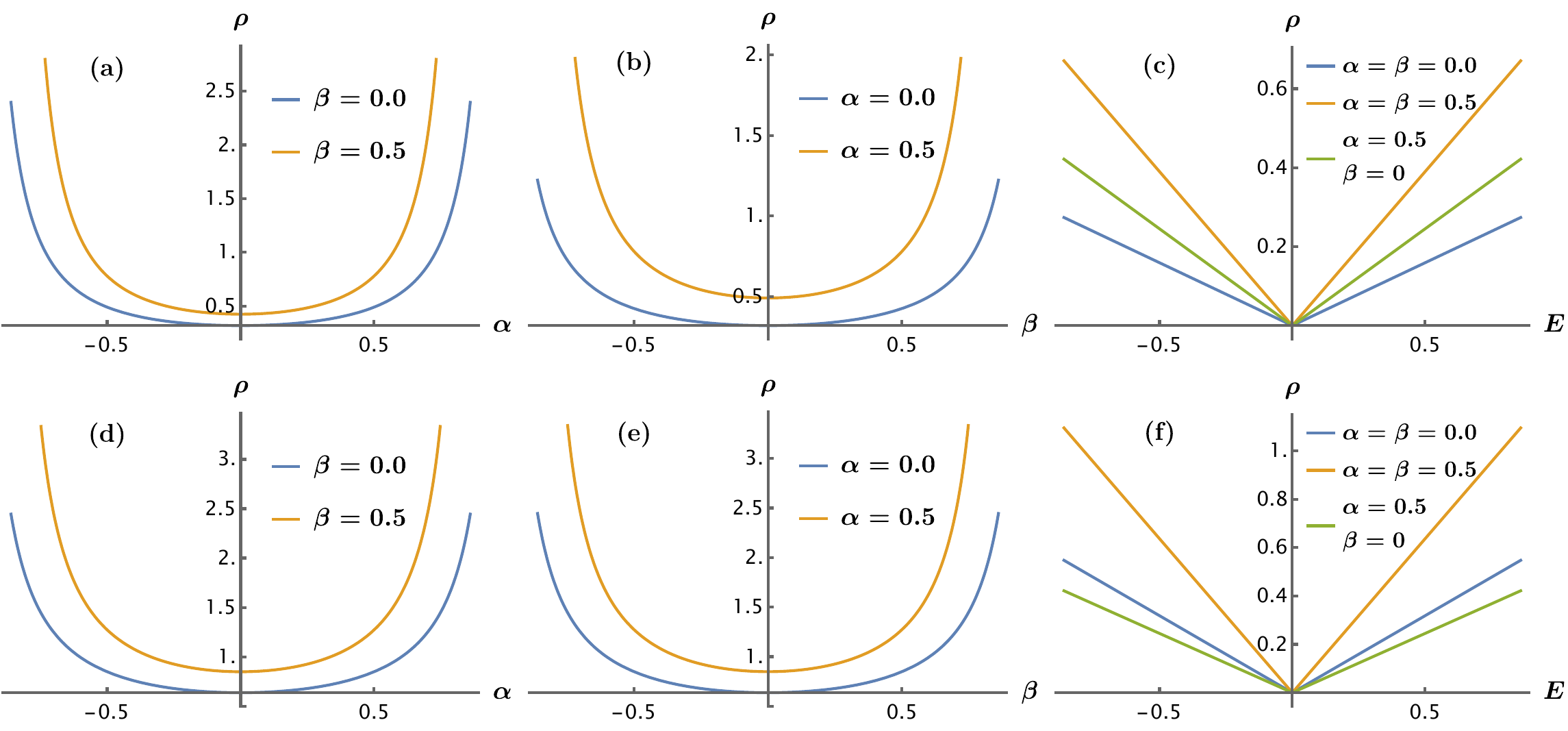}
        \caption{The density of states (DOS) for the antisymmetric tilt [(a), (b) and (c)] and velocity anisotropy case [(d), (e), and (f)], given by Eq.~\eqref{eq:DOS-AS-tilt}   and Eq.~\eqref{eq:DOS-symmetric}, respectively.  (a) and (d). The DOS as a function of the tilt parameter for a fixed non-Hermitian (NH) velocity, $\beta$. (b) and (e). The DOS as a function of the NH parameter $\beta$ for a fixed tilt [panel (b)] and velocity anisotropy [panel (e)]. (c) and (f). The DOS as a function of energy for several values of the parameters $\alpha$ and $\beta$.  In all the plots, we set $\vh=1$. The plots in panels (a), (b), (d) and (e) are displayed for the value of energy $E=1$. }
\label{fig: dos}
    \end{figure*}  

For the non-tilt matrix $T=T_{{\rm VA}}$ in Eq.~\eqref{eq:tilt-Hamiltonian}, the spectrum reads as
\begin{align}
\label{eq:spectrum-velocity-anisotropy}    E_{\textrm{NH}}^{\textrm{VA}}&=\pm  \sqrt{ k^2 \vf^2 \pm  2 \abs{\alpha} \vf \vh k_x^2+\alpha^2 \vh^2 k_x^2}\nonumber\\
&=\pm \vf\sqrt{k_y^2+\left(1\pm\frac{|\alpha|}{\sqrt{1-\beta^2}}\right)k_x^2}\,,
\end{align}
and introduces an effective velocity anisotropy in the Dirac bands since  the matrix $T_{{\rm VA}}$ only partially commutes with the Hermitian Dirac Hamiltonian in Eq.~\eqref{eq:HermitianDirac}. Notice that in this case $[M,T_{{\rm VA}}]=0$, the same as in the case of the tilted NH Dirac Hamiltonian. 

For further considerations, we first display the fermion propagator for the (symmetrically and antisymmetrically) tilted NH Dirac Hamiltonian that  takes the form 
\begin{widetext}
    \begin{align} 
       G^{\textrm{{\rm tilt}}}_F(i\omega,\bfk)=&\Big\{-(i \omega+H_0^{\textrm{t}})A_+ +B(\vh+\vnh M)(\bm{\Gamma}\cdot \bfk)\Tas k_x+2\alpha \vf k_x\big(\vf^2 k^2 \Tas \nonumber\\  
       & +\alpha \vh k_x(\vh+\vnh M)(\bm{\Gamma}\cdot \bfk) \big) \Big\}\frac{1}{A_-^2 -B^2}, 
       \label{eq:GF-asymmetric}
    \end{align}
       where
       \begin{equation}
     A_{\pm}=A_{\pm}(i\omega,k)= \omega^2 +\vf^2 k^2\pm \alpha^2 \vh^2  k_x^2, \quad B=B(i\omega,k)=2i \alpha \vh\omega k_x.
\end{equation}
On the other hand,  for the velocity-antisymmetric NH Dirac Hamiltonian, the propagator reads 

 \begin{align} 
       G^{\textrm{VA}}_F(i\omega,\bfk)&=\Big\{-(i \omega+H_0^{\textrm{t}})A +2i \omega \alpha\vh k_x^2(\vh+\vnh M)T_{{\rm VA}} \Gamma_x\\    &+2\alpha \vh \vf^2 k_x^2(k_x +\Gamma_y \Gamma_x k_y )T_{{\rm VA}}  +2\alpha^2 \vh^2 k_x(\vh+\vnh M)\Gamma_x k_x^3 \Big\}\frac{1}{A^2 -B^2},
       \label{eq:GF-symmetric}
       \end{align}
       where 
\begin{equation}
     A=A(i\omega,\bfk)= \omega^2 +\vf^2 k^2+ \alpha^2 \vh^2  k_x^2, \quad B=B(i\omega,\bfk)=2\alpha \vh\vf k_x^2.
     \end{equation}
See appendix~\ref{a:fermion propagator} for the technical details. As a first step in the analysis of the behavior of the tilted NH Dirac fermions in the presence of the electron-electron interactions, we analyze the DOS. 
\end{widetext}
\subsection{Density of states}
\label{subsec:DoS}
We calculate the DOS in terms of the (Euclidean) fermion Green's functions given as 
\begin{equation}
    \rho(E)= -\frac{1}{\pi}\lim_{\eta\to0^+} \Im\qty{ \Tr G_F(i \omega\to  E+i \eta, \vb{k})}.
    \end{equation}
Using the form of the Green's function for the antisymmetric tilt~\eqref{eq:GF-asymmetric}, we find 
\begin{equation}
\label{eq:DOS-AS-tilt}
 \rho_{\textrm{AS}}(E)= \frac{1}{\pi}\frac{ \vf\abs{E}}{(\vf^2 -\alpha^2 \vh^2)^{3/2}}=\frac{\abs{E}}{\pi (v_F^{\rm tilt})^2}\,\sqrt{\frac{1-\beta^2}{1-\alpha^2-\beta^2}},
 \end{equation}
where the effective tilt-dependent Fermi velocity  $v_F^{\rm tilt}=v_H\sqrt{1-\alpha^2-\beta^2}$. In the following, we only consider the subcritically tilted and weak NH case $|\alpha|^2+|\beta|^2<1$, with pointlike Fermi surfaces and the quasiparticles exhibiting a finite lifetime (type-1). Otherwise, NH Dirac Hamiltonian describes type-2 Weyl/Dirac semimetal~\cite{Soluyanov2015} generalized to the NH case when the tilt parameter $|\alpha|>1$, while the non-Hermiticity is weak, $|\beta|<1$, therefore giving rise to the electron- and hole-like Fermi surfaces with the quasiparticles featuring a finite lifetime. 
For the case of the symmetric tilt, the DOS takes the form  
\begin{equation}
       \rho_{\textrm{S}}(E)=  \frac{\abs{E} }{\pi\vf^2},
       \label{eq:DOS-symmetric-1}
\end{equation}
which is the same as for the non-tilted NH Dirac fermions. Finally, for the velocity-anisotropy [Eq.~\eqref{eq:GF-symmetric}], the DOS reads as 
\begin{equation}
       \rho_{\textrm{VA}}(E)= \frac{1}{\pi} \frac{\abs{E} }{\vf^2-\alpha ^2 \vh^2}=\frac{\abs{E}}{\pi (v_F^{\rm tilt})^2}.
       \label{eq:DOS-symmetric}
\end{equation}
The plots are shown in Fig.~\ref{fig: dos} with the technical details shown in the appendix~\ref{a:DOS}. 

As we can observe, in all the three cases the DOS retains the linear scaling in energy, with the slope depending on the form of the matrix in Eq.~\eqref{eq:NH-Dirac}. 
We notice that for the antisymmetric  tilt, the DOS  explicitly depends on the NH parameter ($\beta$), i.e. it is not solely a function of the effective Fermi velocity, $v_F^{\rm tilt}$. Second, the symmetric tilt acts as an effective chiral chemical potential and therefore the contributions from the two valleys cancel out, yielding the tilt-free form of the DOS, Eq,~\eqref{eq:DOS-symmetric-1}. 
Finally, in case of the velocity anisotropy, the DOS effectively retains the form as in the  tilt-free NH Dirac Hamiltonian with the effective velocity being $v_F^{\rm tilt}$. When the tilt parameter $\alpha=0$, both antisymmetric tilt and velocity anisotropy yield the same tilt-free form of the DOS, given by Eq.~\eqref{eq:DOS-symmetric-1}. 

Such a linear scaling with energy of the DOS, $\rho(E)\sim E$, implies  that the dynamical exponent, measuring the relative scaling between the energy and momentum, is $z=1$. The short-range Hubbard-like electron interaction is therefore  irrelevant for 
$d>1$ because its engineering scaling dimension in units of momentum is $z-d$. In turn,    the tilted  Dirac semimetal remains stable for weak short-range electron-electron interactions. The tendency to a specific ordering at strong coupling then can be deduced from the behavior of the mean-field susceptibility, which we discuss next. 

\begin{figure}[t!]
    \centering
    \includegraphics[scale=0.7]{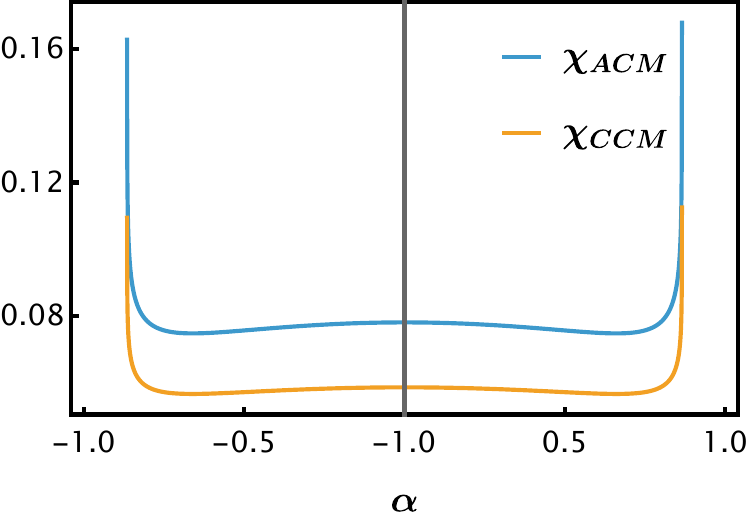}
    \caption{The mean-field susceptibility  for the anticommuting class mass  (ACM) (Eq.~\eqref{eq:susc-tilt-ACM}) and the commuting class mass (CCM)  (Eq.~\eqref{eq:susc-tilt-CCM}) orders  when noninteracting non-Hermitian (NH)  Dirac Hamiltonian features antisymmetric tilt term. We notice that  peaks in the susceptibility emerge  for the tilt $\abs{\alpha}= \sqrt{1-\beta^2}$ corresponding to the divergence of the density of states in Eq.~\eqref{eq:DOS-AS-tilt} at the zero effective velocity $v_F^{\rm tilt}$. The values of Hermitian and NH velocities are  $\vh=1$ and $\vnh=0.5$, respectively, which corresponds to the  values of  the non-Hermiticity parameter $\beta=1/2$ and the effective Fermi velocity  $\vf=\sqrt{3}/2$.  }
    \label{susc as}
\end{figure}

\section{Mean-field susceptibilities for the mass orders}
\label{sec:susceptibilities}
The OP that isotropically gaps out hermitian Dirac fermions can be expressed as an $n$-component vector $\Phi_j=\langle\Psi^\dag N_j \Psi\rangle$ such that $\{H_{\textrm{D}},N_j\}=0$. Here, the $(4N_f\times4N_f)$ matrices $N_j$, $j=1,..,n$,   mutually anticommute, 
$\{N_j,N_k\}=2\delta_{jk}$, with $N_f$ as the number of four-component Dirac fermion flavors. The matrix $M$ in the NH Dirac Hamiltonian induces fragmentation of the possible OPs and we here  analyze two cases: $(i)$ A CCM OP for which $[N_j,M]=0$, for $j=1\,,\,2\,,\,...\,,n$;  (ii) An ACM OP for which $\{N_j,M\}=0$ for $j=1\,,\,2\,,\,...\,,n$ . We do not consider a mixed class where the OP matrix partially (anti)commutes with the matrix $M$. 

\begin{figure}[t!]
    \centering
    \includegraphics[scale=0.65]{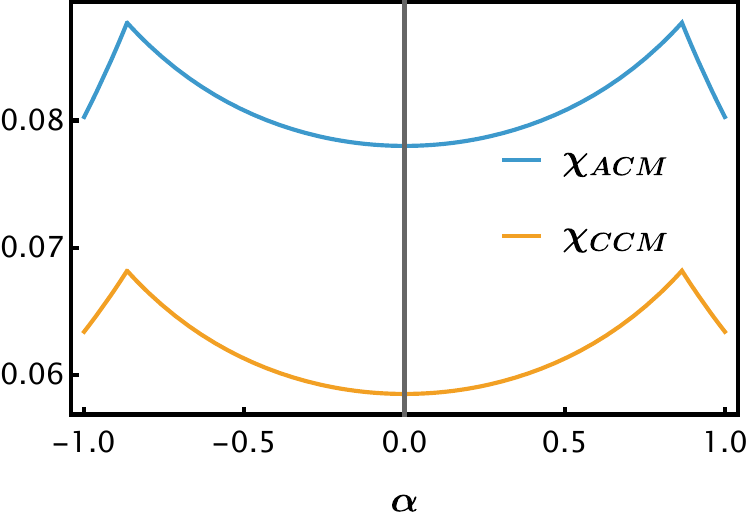}
    \caption{The mean-field susceptibility  for the anticommuting class mass (ACM) [Eq.~\eqref{eq:susc-VA-ACM}] and commuting class mass (CCM)  [Eq.~\eqref{eq:susc-VA-CCM}] orders  when noninteracting non-Hermitian Dirac Hamiltonian features velocity-anisotropy term.   We notice two cusps corresponding to the absolute value of the velocity-anisotropy parameter  $\abs{\alpha}= \sqrt{1-\beta^2}$, with $\beta=\vnh/\vh$. These cusps coincide with the maximum slope of the density of states, as given by  Eq.~\eqref{eq:DOS-symmetric},  with the effective velocity in 
 Eq.~\eqref{eq:spectrum-velocity-anisotropy}, and therefore indicate increased propensity toward  the instabilities therein. The values of the parameters are the same as in Fig.~\ref{susc as}.   }
    \label{susc s}
\end{figure}

The ordering tendencies for these two classes of instabilities, showing the influence of the NH parameter ($\beta$), can be estimated  from their respective (nonuniversal) bare mean-field susceptibilities at zero external frequency and momentum, which for an OP represented by a fermion bilinear $\langle\Psi^\dag N \Psi\rangle$  reads
\begin{equation}
    \chi_N=-\Tr \int_{-\infty}^\infty \frac{d \omega}{2 \pi} \int \frac{d^d \vb{k}}{(2\pi)^d}[N\, G_F(i \omega ,\vb{k})\,N\, G_F( i \omega ,\vb{k})], 
\end{equation}
since the (nonuniversal)  critical interaction strength for such an ordering $g_{N,*}\sim \chi_N^{-1}$.
Here,  $ G_F(i \omega ,\vb{k})$ is the fermion propagator for the corresponding noninteracting Hamiltonian and we assume an upper cutoff $\Lambda$ in the momentum integral, and consider $d=3$  being the upper critical dimension for the corresponding quantum-critical theory (Sec.~\ref{sec:NH-critical}). 
We furthermore consider the susceptibilities  in the two cases with qualitatively different spectrum in the noninteracting NH Dirac Hamiltonian~\eqref{eq:tilt-Hamiltonian}, namely, the antisymmetric tilt and the velocity  anisotropy, while the symmetric tilt is not explicitly considered in the following  since it is  analogous to its antisymmetric counterpart.

\begin{widetext}
First, in the case of the antisymmetric tilt, the susceptibility corresponding to an ACM order has the form 
\begin{equation}
        \chi_N^{\textrm{ACM}}= h_{\textrm{AS}}(\vf,\beta,\alpha)\Lambda^2,
        \label{eq:susc-tilt-ACM}
\end{equation}
where
\begin{align}
     &h_{\textrm{AS}}(\vf,\beta,\alpha)=\frac{1}{32 \pi^3\vf}\int [d\phi d\theta] \frac{  (1-\beta^2) \left(-2 \alpha ^2 \sin ^2\theta  \cos ^2\phi +\alpha ^2+4\right)-2 \alpha ^2 }{(1-\beta^2)(1-\beta^2-\alpha ^2 \sin ^2\theta  \cos ^2\phi) }\nonumber\\
    &=\frac{ \left(1+\beta ^2\right) \left(2-\alpha ^2-2 \beta ^2\right) \tanh ^{-1}\left(\frac{\alpha }{\sqrt{1-\beta ^2}}\right)+2 \alpha  \left(1-\beta ^2\right)^{3/2}}{8 \pi ^2 \alpha  \left(1-\beta ^2\right)^{3/2} \vf}\,,
\end{align}
with $[d\phi d\theta]\equiv \int_0^{2\pi}d\phi \int_0^{\pi}d\theta\sin\theta$.

For the CCM order, the susceptibility takes the form 
\begin{equation}
    \chi_N^{\textrm{CCM}}= f_{\textrm{AS}}(\vf,\beta,\alpha) \Lambda^2,
    \label{eq:susc-tilt-CCM}
\end{equation} 
where 
\begin{align}
    f_{\textrm{AS}}(\vf,\beta,\alpha)&=\frac{1}{16 \pi^3 \vf}\int [d\phi d\theta]\,  \frac{  -\alpha ^2  \left(2 \sin ^2\theta  \cos ^2\phi +1\right)+4(1-\beta^2)}{2 \left(1-\beta^2-\alpha ^2  \sin ^2\theta   \cos ^2\phi \right)}
    =\frac{1 }{16 \pi^2 \vf\sqrt{1-\beta ^2}}\Bigg[  \alpha  \ln \left(\frac{\sqrt{1-\beta ^2}-\alpha }{\alpha +\sqrt{1-\beta ^2}}\right)\nonumber\\
    &+4 \sqrt{1-\beta ^2}+\frac{2}{\alpha} (\beta ^2-1) \ln \left(\frac{\sqrt{1-\beta ^2}-\alpha }{\alpha +\sqrt{1-\beta ^2}}\right)\Bigg].
\end{align}

\end{widetext}

Both functions also satisfy the scaling as $\sim \vf^{-1}$, with their explicit form for ACM and CCM OPs shown in Fig.~\ref{susc as}. We notice that the susceptibility is larger for the ACM than for the CCM case, implying the lower critical coupling in the former case. Therefore, the ACM order is favored in the case of antisymmetric tilt, as expected from the fact that the ACM order opens a gap at the band touching points in the tilted Dirac semimetal. Finally, the cusps in the susceptibilities in Fig.~\ref{susc as} are a consequence of the vanishing effective Fermi velocities at these specific values of the tilt parameter which makes the system more prone to the instabilities due to the ensuing divergent DOS.

In  the velocity anisotropic semimetal, for the CCM case, we obtain 
\begin{align}
\label{eq:susc-VA-CCM}
    \chi_N^{\textrm{CCM}}&= f_{\textrm{VA}}(\vf,\beta,\alpha) \Lambda^2\,,
\end{align} 
where
\begin{widetext}
\begin{equation}
    f_{\textrm{VA}}(\vf,\beta,\alpha)=
   \frac{1}{16\pi^3 \vf}\sum_{\zeta=\pm}\int [d\phi d\theta]\,\, \sqrt{\frac{1-\beta^2}{\alpha  \left(2 \zeta\sqrt{1-\beta ^2}+\alpha \right) \sin ^2(\theta ) \cos ^2(\phi )+1-\beta ^2}}.
\end{equation}
\end{widetext}

Now, for the ACM case, we find  the susceptibility of the form 
\begin{equation}
\label{eq:susc-VA-ACM}
  \chi_N^{\textrm{ACM}}= \qty( f_{\textrm{VA}}(\vf,\beta,\alpha)+h_{\textrm{VA}}(\vf,\beta,\alpha))\Lambda^2\,,  
\end{equation}
where
\begin{align}
   & h_{\textrm{VA}}(\vf,\beta,\alpha)=
   -\frac{1}{16 \pi^3} \frac{\beta ^2 }{\vf\alpha  }\sum_{\zeta=\pm}\int [d\phi d\theta]\csc^2 (\theta ) \sec ^2(\phi )\nonumber\\
   &\times \zeta\sqrt{\frac{1}{\alpha  \left(2 \zeta\sqrt{1-\beta ^2}-\alpha \right) \sin ^2(\theta ) \cos ^2(\phi )+1-\beta ^2}}.
\end{align}
We notice that the susceptibilities for both CCM and ACM OPs scale as $\sim \vf^{-1}$. Furthermore, the ACM order is also favored in the case of velocity-anisotropic NH Dirac semimetal, as explicitly shown in Fig.~\ref{susc s}. Notice that susceptibilities for both ACM and CCM orders are increased in comparison to the tilt-free and isotropic semimetal, with a particularly pronounced cusps when the value of the effective velocity reaches zero, and therefore the DOS diverges.

The susceptibilities for both antisymmetrically tilted and velocity-anisotropic NH Dirac semimetals  in $d=2$ are  analyzed in Appendix~\ref{a:susceptibilities-d2}, and show  qualitatively similar behavior as in $d=3$. Particularly, the ACM order is favored in both cases. However, to address the competition between the different instabilities, together with the emergence of the Yukawa-Lorentz symmetry, we need to consider the RG flows of the velocity parameters and of the tilt in the quantum-critical regime. This is even more pertinent given a rather small difference between the mean-field susceptibilities for the two classes of the OPs, for both the asymmetric tilt (Figs.~\ref{susc as}) and velocity anisotropy (Fig.~\ref{susc s}). Motivated by this, we now analyze the quantum-critical behavior of NH Dirac semimetals with the antisymmetric tilt and velocity anisotropy in the vicinity of the quantum phase  transition corresponding to the ACM and CCM OPs condensation. 
\section{Quantum-critical theory of the tilted non-Hermitian Dirac fermions}
\label{sec:NH-critical}
To address the quantum-critical behavior of the tilted NH Dirac fermions, we employ the zero-temperature ($T=0$) Euclidean  action  for $d-$dimensional  tilted NH Dirac fermions 
\begin{equation}
\label{accion fermionica}
    S_F[\Psi,\Psi^\dagger]=\int\, d\tau\, d^d{\vb{r}}\, \Psi^\dagger \qty[\partial_\tau+ H_{\textrm{NH}}^{\textrm{tilt}}(\vb{k}\to -i \nabla_{\vb{r}})]\Psi\,,
\end{equation}
where $\tau$ is the imaginary time, and $\Psi\equiv\Psi(\tau,{\bf r})$.

The bosonic OP fluctuations are described by a $\Phi^4$ theory, with a bosonic action of the form,
\begin{widetext}
\begin{equation}
\label{accion bosonica}
    S_B[\Phi]=\int d\tau d^d{\vb{r}}\qty[\sum_{j=1}^n\Phi_j(-\partial_\tau^2-\vB^2\nabla^2+m^2_B)\Phi_j  + \frac{\lambda}{2}\qty(\sum_{j=1}^n\Phi^2_j)^2]\,,
\end{equation}
\end{widetext}
where a bosonic propagator $G_B(i \omega,\vb{k})=[\omega^2 +\vB^2 k^2+m^2_B ]^{-1}$. Since we consider  the critical plane of the theory, we set $m_B=0$.

  \begin{figure*}[t!]
        \centering      \includegraphics[scale=1]{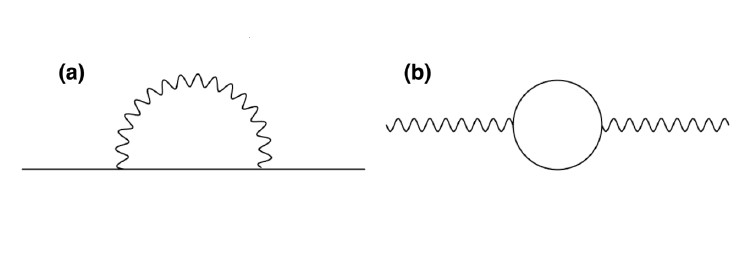}
        \caption{Self-energy diagrams for the Dirac fermion (solid lines) in panel (a) and the bosonic order parameter field (wavy lines) in panel (b). The vertex in both diagrams 
corresponds to the Yukawa coupling in Eq.~\eqref{gny}. }
\label{fig:self-energy}
    \end{figure*}  

 The coupling of the bosonic fluctuations to the NH tilted fermions is described  by a Yukawa interaction of the form
\begin{equation}
    S_Y=g \int d\tau \int d^d \vb{r}\sum_{j=1}^n \Phi_j(\tau,\vb{r}) \Psi^\dag(\tau,\vb{r}) N_j\Psi(\tau,\vb{r}) \,, \label{gny}
\end{equation}
which encodes the effects of short-range Coulomb interaction in the quantum-critical region, where the fermionic and the bosonic fluctuations are strongly coupled.
The Yukawa, and the $\Phi^4$ couplings, $g$ and $\lambda$, respectively, are both marginal in $d=3$, the  upper critical dimension of the theory,  since their engineering dimensions are $\textrm{dim}[g^2]=\textrm{dim}[\lambda]=3-d$, with the engineering scaling dimension of momentum set to unity.  We thus use the distance from the upper critical dimension $\epsilon=3-d$ as expansion parameter to analyze the quantum-critical theory in $d=2$~\cite{ZinnJustin2002,Peskin2019}. To this end, we define a Euclidean renormalized action, taking  Eqs.~\eqref{accion fermionica}-\eqref{gny}, which reads 
\begin{widetext}
\begin{align}
   S_R=&\int d\tau d^d{\vb{r}}\Bigg[Z_{\Psi}\Psi^\dagger [\partial_{\tau}-i\qty(Z_{\vh}  \vh  +Z_{\vnh}\vnh M)\qty(\Gamma_x\partial_x +\Gamma_y\partial_y)-i Z_{\alpha} Z_{\vh}\alpha\vh T\partial_x]\Psi\nonumber\\& +Z_{\Phi}\qty(\sum_{j=1}^n\Phi_j(-\partial_\tau^2-Z_{\vB^2}\vB^2\nabla^2)\Phi_j )+ Z_{\lambda}\frac{\lambda}{2}\qty(\sum_{j=1}^n\Phi^2_j)^2+Z_g g\sum_{j=1}^n \Phi_j(\tau,\vb{r}) \Psi^\dag(\tau,\vb{r}) N_j\Psi(\tau,\vb{r})\Bigg]\,,
   \label{eq:SR-full}
\end{align}
\end{widetext}
where  $Z_{Q}$ are renormalization constants. Furthermore, we define renormalization  conditions for each parameter  relating  the renormalized quantities with the bare ones at the cutoff scale, 
\begin{align}
    Z_{\vh}\vh&=\vh^0\\
    Z_{\vnh}\vnh &=\vnh^0\\
    Z_{\alpha}\alpha&=\alpha^0\\
    Z_{\vB}\vB^2&=(\vB^0)^2
\end{align}
where the quantities with (without) index $0$ in the superscript are bare (renormalized).

To analyze the fate of the Yukawa-Lorentz symmetry at the quantum criticality,  we compute the fermionic [Fig.~\ref{fig:self-energy}(a)] and bosonic [Fig.~\ref{fig:self-energy}(b)]  self-energy diagrams to the leading order in the $\epsilon$ expansion.
After computing the self-energy diagrams, as shown in  Appendix \ref{a:renormalization}, we find  the renormalization factors $Z_{Q}$ to the one-loop order in the minimal-subtraction-scheme. Using the obtained factors,  we subsequently compute the RG $\beta$ functions, which are defined as $\beta_Q\equiv d Q/ d \ln b|_{Q^0}$, with the bare quantities fixed, and $b$ as the RG scale, rescaling the UV cutoff in the theory, $\Lambda\to\Lambda/b$, in  the RG transformation. These RG $\beta$ functions take the form
\begin{align}
\label{eq:beta-function-vh}
     \beta_{\vh} &= -\vh n \frac{g^2}{2}\qty[G(\alpha, \vh,\vnh, \vB)-J(\alpha, \vh,\vnh, \vB)] \,,\\
     \label{eq:beta-function-vnh}
      \beta_{\vnh} &= -\vnh n\frac{g^2}{2}\qty[G(\alpha, \vh,\vnh, \vB)\mp J(\alpha, \vh,\vnh, \vB)] \,,\\
       \label{eq:beta-function-alpha}
       \beta_{\alpha} &= -\alpha n\frac{g^2}{2}\qty[J(\alpha, \vh,\vnh, \vB)-I(\alpha, \vh,\vnh, \vB)] \,,\\
        \label{eq:beta-function-vb}
      \beta_{\vB} &= -\vB n N_f g^2\qty[F(\alpha, \vh,\vnh, \vB)-C(\alpha, \vh,\vnh, \vB )] \,,
\end{align}
where we rescaled the Yukawa coupling, $g^2/(8\pi^2)\to g^2$, and $G\,, I\,,J\,,F\,,C$ are functions defined  in Appendix  $\ref{a:renormalization}$. The upper (lower) sign in Eq.~\eqref{eq:beta-function-vnh} corresponds to the CCM (ACM) order. We here do not discuss the flow of the Yukawa and the $\lambda$ couplings for the following reasons. First, the RG flow of the Yukawa coupling  is decoupled from that of $\lambda$ in the leading order of the $\epsilon-$expansion, as is the case for the untilted NH Dirac fermions~\cite{jurivcic2024yukawa}. As we are only interested in the fate of the NH terms and ultimately Lorentz symmetry  at the criticality, the precise value of $g$ at the QCP is unimportant, as can be directly inferred from the RG flow equations in Eqs.~\eqref{eq:beta-function-vh}-\eqref{eq:beta-function-vb}. Second, since the $\lambda$ coupling does not enter the RG flow of either fermionic or bosonic velocities, its detailed behavior at QCP is unimportant as far as the restoration of Lorentz symmetry is concerned. 

\begin{figure*}[t!]
        \centering
       \includegraphics[scale=0.53]{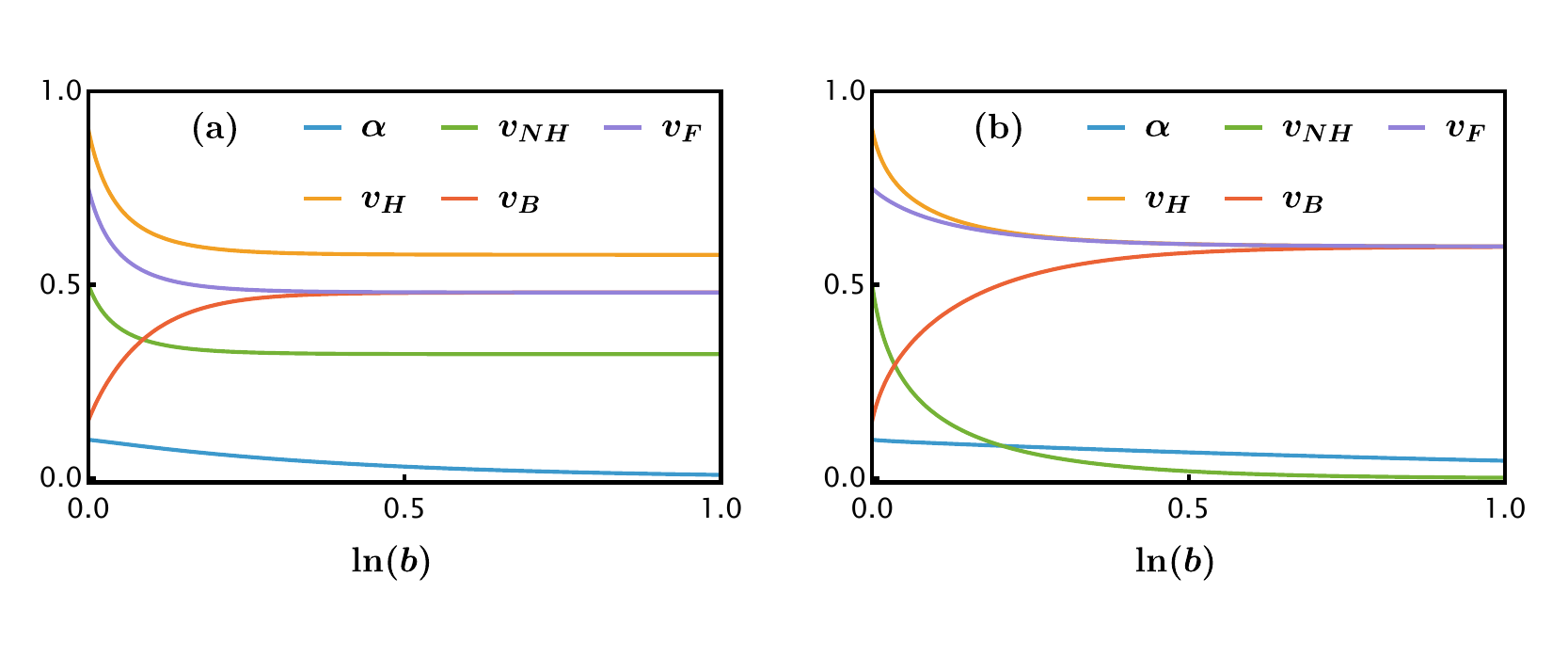}
        \caption{Renormalization-group (RG)  flow of the velocities in the critical plane ($m_B=0$) of the Gross-Neveu-Yukawa theory with antisymmetric tilt and the corresponding RG flow equations given by Eqs.~\eqref{eq:beta-function-vh}-\eqref{eq:beta-function-vb}. (a) Commuting class mass order parameter. (b) Anticommuting class mass order parameter. In both cases we set the bare values of the  tilt and velocity parameters $\alpha^0=0.1,\, \vh^0=0.9,\, \vnh^0=0.5,\, \vB^0=0.15 $, the Yukawa coupling $g=1$, and the number of the order-parameter components $n=1$. }
\label{fig: assym}
    \end{figure*}

To analyze the behavior of the velocities and tilt under the RG at the criticality, we plot the corresponding RG flows for a fixed value of the Yukawa coupling, $g$. Ultimately, the form of the RG flows does not depend on the value of this coupling, as also Eqs.~\eqref{eq:beta-function-vh}-\eqref{eq:beta-function-vb} explicitly show. The corresponding RG flows  for the antisymmetric tilt are displayed in the Fig.~\ref{fig: assym}, and for the velocity-antisymmetric NH Dirac fermions   in Fig.~\ref{fig: symm}, with  CCM or ACM OPs considered in both cases. 

We first notice that for both antisymmetric tilt and velocity anisotropy, the qualitative features of the RG flow are similar. Both the tilt and the velocity anisotropy flow to zero under the RG transformation, i.e. they turn irrelevant in the quantum-critical region. The rate of the RG flow, as it can be noticed, is slightly faster for the CCM than for the ACM OP in the case of the tilt. Furthermore, for the CCM OP in the case of either tilt or velocity anisotropy, the NH velocity flows to a finite value, while the terminal bosonic velocity and the effective Fermi velocity reaches a common value.  Therefore the NH Yukawa-Lorentz symmetry is restored in spite of the symmetry-breaking tilt or velocity anisotropy present in the bare (lattice) theory. In the case of ACM order, the NH velocity parameter vanishes, irrespective of the additional (tilt or velocity anisotropy) term, and therefore the system spontaneously restores hermiticity, and at the same time, conventional Lorentz symmetry.

\section{Conclusions}
\label{sec:conclusions}

In summary, our analysis of quantum-critical NH tilted Dirac fermions shows that Yukawa-Lorentz symmetry emerges  near the strongly coupled QCP separating the NH Dirac semimetallic and insulating phases.  This key feature arises from the irrelevance of the tilt term near the QCP, as our leading-order renormalization group calculations explicitly demonstrate, see Eqs.~\eqref{eq:beta-function-vh}-\eqref{eq:beta-function-alpha}.  Crucially, the fate of non-Hermiticity hinges on the OP class (ACM or CCM), with ACM leading to complete decoupling from the environment and Hermiticity restoration, while CCM maintains coupling to the environment yet exhibits emergent Yukawa-Lorentz symmetry, see also Fig.~\ref{fig: assym}. Furthermore,  a velocity anisotropy term turns out to be   irrelevant near the QCP (Fig.~\ref{fig: symm}).  These results strongly suggest that Yukawa-Lorentz symmetry is a universal feature of quantum-critical NH Dirac fermions, a prediction that can be tested  through numerical lattice  simulations. 

For the quantum phase transition from NH tilted Dirac semimetal to an ACM OP, even though, the tilt eventually vanishes, its effect on the observables should be non-negligible, due to the finite system size. On the other hand, in the case of the CCM OP, non-Hermiticity close to the QCP should have observable  effects on the correlators through a contribution of the finite NH velocity, which we plan to study in near future. 

We would like to emphasize  that the NH piece of the Dirac Hamiltonian directly corresponds to the nonreciprocity of the nearest-neighbor hopping on the honeycomb lattice, while the tilt term can be introduced  by the uniaxial strain along the zigzag direction, for instance~\cite{Goerbig-PRB2008}. However, a remaining open question concerns  the precise emergence of these NH terms from coupling to an excitation bath with a specific spectral function. Establishing such a correspondence will corroborate engineering of the NH terms and   deserves further investigation.

Finally, we point out that our results should motivate further studies of the NH tilted Dirac fermions, as for instance, their instabilities in the case when they feature a finite-size 
Fermi surface with finite-lifetime quasiparticles (NH type-2 Dirac fermions). The studies of the NH tilted Dirac fermions may also provide some valuable insights regarding the relationship between the condensed matter and low-dimensional gravitational systems, particularly concerning the effects of the Hawking radiation that may be captured through the effective NH Dirac operators discussed here.

\acknowledgments 

The authors are grateful to Bitan Roy for the critical reading of the manuscript.  This work was supported by Fondecyt (Chile) Grant No.  1230933 (V.J.), the Swedish Research Council Grant No. VR 2019-04735 (V.J.), and Anillo Grant ANID/ACT210100 (V.J.).

\appendix
\begin{widetext}
\section{Minimal model}

In this appendix, we provide the details of the minimal model, given by the Hamiltonian in Eq.~\eqref{eq:tilt-Hamiltonian}, for the NH tilted Dirac fermions. 


\subsection{Fermionic propagator}
\label{a:fermion propagator}

We can obtain our fermion propagator noting that the inverse of the propagator is
\begin{equation}
 G_F^{-1}(i\omega, \bfk )={} i\omega-\qty[(\vh+\vnh M)\qty(\Gamma_x k_x + \Gamma_y k_y)+\alpha \vf T k_x].
\end{equation}

In the case of antisymmetric tilt, we obtain 
\begin{equation}
     (G_F^{\textrm{AS}}(i\omega,\bfk))^{-1}(i\omega+\qty[(\vh+\vnh M)(\bm{\Gamma}\cdot \bfk)+\alpha \vh \Tas k_x])=-\omega^2- \vf^2k^2-\alpha^2\vh^2k_x^2\\ 
     -2\alpha\vh(\vh+\vnh M)(\bm{\Gamma}\cdot \bfk)\Tas k_x. \nonumber 
     \end{equation}
     
    Again, we carry out a multiplication to get a pure scalar function (proportional to the unity matrix) on the right hand side with $\Gamma$ or $T$ matrices,
   \begin{align}
    & (G_F^{\textrm{AS}}(i\omega,\bfk))^{-1}\qty(i\omega+\qty[(\vh+\vnh M)(\bm{\Gamma}\cdot \bfk)+\alpha \vh \Tas k_x]) \times (-\omega^2- \vf ^2k^2-\alpha^2\vh^2k_x^2+2\alpha\vh(\vh+\vnh M)(\bm{\Gamma}\cdot \bfk)\Tas k_x\nonumber\\
    &=(\omega^2+\vf^2k^2+\alpha^2 \vh^2 k_x^2)^2-4\alpha^2 \vf^2 \vh^2 k_x^2. 
     \end{align}
     Finally, we multiply by the Green's function from the left to find 
     \begin{align}  G^{\textrm{AS}}_F(i\omega,\bfk)&=\Big\{-(i \omega+H_0^{\textrm{tilt}})A_+ +B(\vh+\vnh M)(\bm{\Gamma}\cdot \bfk)\Tas k_x \\  &\quad +2\alpha \vf k_x(\vf^2 k^2 \Tas +\alpha \vh k_x(\vh+\vnh M)(\bm{\Gamma}\cdot \bfk) ) \Big\}\frac{1}{A_-^2 -B^2}, \nonumber \\
       \intertext{with}
     A_{\pm}&=A_{\pm}(i\omega,\bfk)= \omega^2 +\vf^2 k^2\pm \alpha^2 \vh^2  k_x^2\,, \quad B=B(i\omega,\bfk)=2i \alpha \vh\omega k_x \,.
     \label{eq:Aplusminus}
\end{align}

 \begin{figure*}[t!]
        \centering
    \includegraphics[scale=0.53]{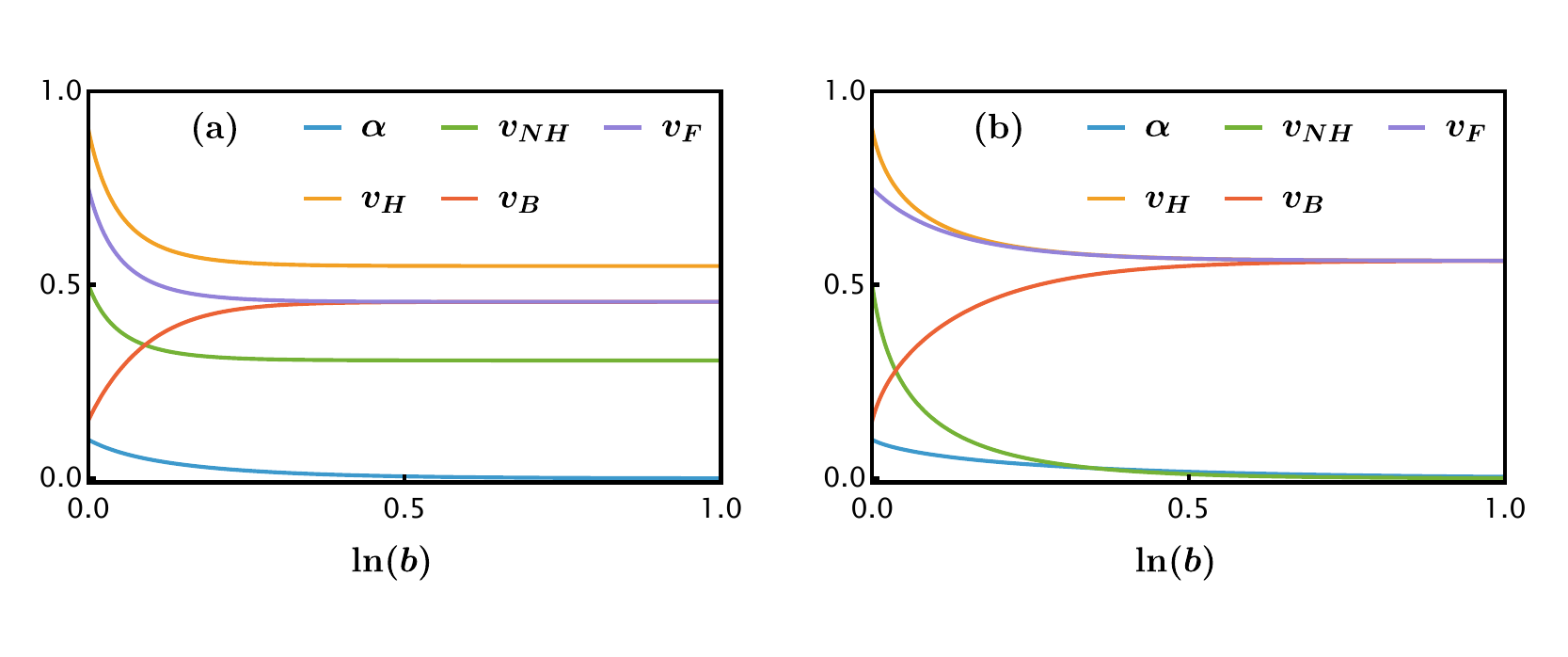}
        \caption{Renormalization-group (RG) flow of the velocity parameters in the critical plane ($m_B=0$) of the
Gross-Neveu-Yukawa theory for  the quantum-critical NH Dirac fermions, given by Eqs.~\eqref{accion fermionica}-\eqref{gny},  with velocity anisotropy. (a) Commuting class mass order parameter.   (b) Anticommuting mass order parameter. We use the same  values of the parameters as in Fig.~\ref{fig: assym}.}
      \label{fig: symm}
    \end{figure*}  

In the case of the velocity anisotropy,   carrying out the same procedure, and  using  that $[\Gamma_x,\Ts]=0$, $\{\Gamma_y,\Ts\}=0$, we obtain the Green's function in the form 
\begin{align} 
G^{\textrm{S}}_F(i\omega,\bfk)&=\Big\{-(i \omega+H_0^{\textrm{tilt}})A_+ +2i \omega \alpha\vh k_x^2(\vh+\vnh M)\Ts \Gamma_x\\\quad  &+2\alpha \vh \vf^2 k_x^2(k_x +\Gamma_y \Gamma_x k_y )\Ts   +2\alpha^2 \vh^2 k_x(\vh+\vnh M)\Gamma_x k_x^3 \Big\}\frac{1}{A^2 -B^2}, \nonumber\\
       \intertext{where we defined}
     A&=A(i\omega,\bfk)= \omega^2 +\vf^2 k^2+ \alpha^2 \vh^2  k_x^2\,, \quad B=B(i\omega,\bfk)=2\alpha \vh\vf k_x^2 \,. 
\end{align}
\subsection{Calculation of the density of states}
\label{a:DOS}

To calculate the DOS in terms of the Green's function, we use 
\begin{equation}
    \rho(E)= -\frac{1}{\pi}\lim_{\eta\to0^+} \Im\qty{ \Tr G_F(i \omega\to E+i \eta, \vb{k})}.
    \end{equation}

After taking the trace of the Green's function, we use its partial fraction decomposition.  
We then apply the analytical continuation, $i\omega\to\omega+i\eta$, and finally use the Sokhotski-Plemelj theorem for the real line, 
\begin{equation}
    \lim_{\eta \to 0^{+}} \frac{1}{x\pm i \eta }= \mp i \pi \delta(x) + \mathcal{P} \qty(\frac{1}{x}),
\end{equation}
where $\mathcal{P}$ denotes the principal value, while $\delta(x)$ is the delta function. 

Now, carrying out this procedure,  we calculate the DOS, which for the antisymmetric tilt yields, 
\begin{align}
    \rho_{\textrm{AS}}(E)=& -\frac{1}{\pi}\lim_{\eta\to0^+} \Im\qty{ \Tr G_F^{\textrm{AS}}(i \omega\to E+i \eta, \vb{k})}\\
   =& 2\sum_{\zeta=\pm}\int \frac{d^2 \vb{k}}{(2 \pi )^2} \,\delta[E- (\zeta\vf k + \alpha \vh k_x)]\,,\\
    =&  \frac{1}{2 \pi ^2}\sum_{\zeta=\pm} \int_0^{2\pi}d\phi \int_0^\infty dk\, k\, \delta[E- (\zeta k \vf + \alpha \vh k \cos \phi )]\,,
\end{align}
with the argument of each delta function representing the spectrum of a  band of the  Hamiltonian in Eq.~\eqref{eq:spectrum-asymmetric}. 
The DOS can then be compactly written as 
\begin{align}
   & \rho_{\textrm{AS}}(E)= \frac{1}{\pi } \frac{ \vf\abs{E}}{\qty(\vf^2-\alpha^2 \vh^2)^{3/2}}= \frac{\sqrt{1-\beta^2}\abs{E}}{\pi (1-\alpha^2 -\beta^2)^{3/2} \vh^2}\,,
 \end{align}
 where $\beta= \vnh/\vh$, and $|\alpha|<1$, $|\beta|<1$. This is the form of the DOS for the antisymmetric tilt, given by Eq.~\eqref{eq:DOS-AS-tilt}. 

 The same steps  as for the antisymmetric  case yield the DOS for the symmetric tilt and velocity anisotropic  model  as given in Eq.~\eqref{eq:DOS-symmetric-1}  and Eq.~\eqref{eq:DOS-symmetric}, respectively.

\section{Mean-field susceptibilities in $d=2$ }
\label{a:susceptibilities-d2}
In Sec.~\ref{sec:susceptibilities}, we have presented the calculation of the mean-field susceptibilities in the upper critical dimension of the theory, $d=3$. For completeness, we here carry out the mean-field susceptibility calculations directly in $d=2$, which for the OP represented by the fermion bilinear $\langle\Psi^\dag N \Psi\rangle$ reads 

\begin{equation}
    \chi_N=-\Tr \int_{-\infty}^\infty \frac{d \omega}{2 \pi} \int \frac{d^2 \vb{k}}{(2\pi)^2}[N\, G_F(i \omega ,\vb{k})\,N\, G_F( i \omega ,\vb{k})], 
\end{equation}
with $G_F( i \omega ,\vb{k})$ as the fermion Green's function. 

\begin{figure}[t!]
    \centering
    \includegraphics[scale=0.6]{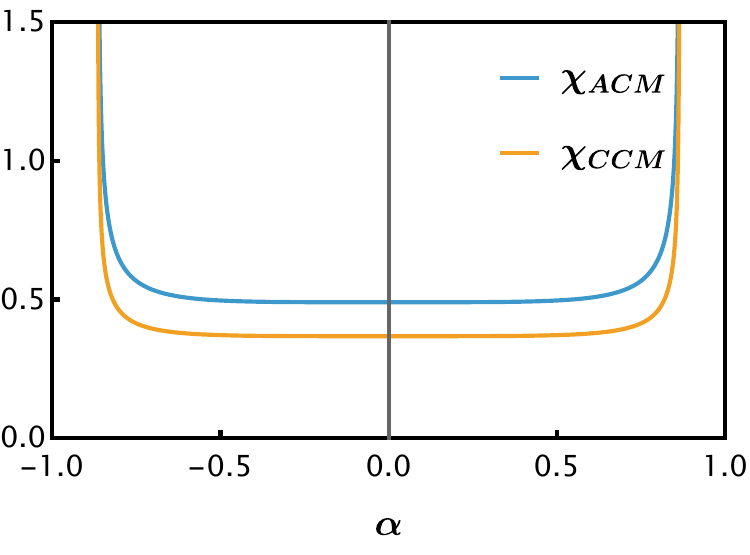}
    \caption{Mean field susceptibility of the  noninteracting non-Hermitian (NH) Dirac Hamiltonian in $d=2$ with the  antisymmetric tilt term, corresponding to the commuting (anticommuting) class mass, given by Eq.~\eqref{eq:susc2dccm} (Eq.~\eqref{eq:susc2dacm}). We notice that the susceptibility diverges when 
    $\abs{\alpha}= \sqrt{1-\beta^2}$, corresponding to the exceptional point of the NH Hamiltonian. We use the  values of  
    NH parameter $\beta=1/2$ and the effective Fermi velocity  $\vf=\sqrt{3}/2$.}
    \label{susc 2d as}
\end{figure}

We first consider the antiymmetric tilt case with the OP matrix $N$ belonging to the CCM class, $[M,N]=0$, 

\begin{align}
\label{eq:susc2dccm}
\chi_N^{\textrm{CCM}}&=
\left(\frac{\Lambda}{4\pi\vf}\right)
\left(\frac{2 \sqrt{(1-\beta ^2)(1-\alpha ^2-\beta ^2)}+2-\alpha ^2-2 \beta ^2}{  \sqrt{(1-\beta ^2)(1-\alpha ^2-\beta ^2)} }\right),
\end{align}
while in the case of an ACM OP, we find 
 \begin{align}  
 \label{eq:susc2dacm}
\chi_N^{\textrm{ACM}}&=\left(\frac{\Lambda}{4\pi\vf}\right) \left( \frac{2 \left(1-\beta ^2\right) \left(\sqrt{(1-\beta ^2)(1-\alpha ^2-\beta ^2)}+\beta ^2+1\right)-\alpha ^2 \left(\beta ^2+1\right)}{  {\left(1-\beta ^2\right)^{3/2} \sqrt{1-\alpha ^2-\beta ^2} }}\right).
\end{align}
Behavior of the susceptibilities for the classes of the mass order in $d=2$, as shown in Fig.~\ref{susc 2d as}, is qualitatively  similar to the situation in $d=3$, namely, $\chi_N^{\textrm{ACM}}>\chi_N^{\textrm{CCM}}$, with  the nucleation of the ACM order being more  favorable than of the CCM class. 

When the system features velocity anisotropy, as given by Eq.~\eqref{eq:tilt-Hamiltonian} with the matrix  $T=T_{{\rm VA}}=\sigma_1\otimes\tau_0$, the susceptibility for the CCM OP reads
\begin{align}
\chi_N^{\textrm{CCM}}&=\frac{\Lambda \sqrt{1-\beta^2}}{(2\pi)^2\vf}\sum_{\zeta=\pm}\int_0^{2\pi}d\phi \, \,\frac{1}{\sqrt{\alpha  \left(\alpha +2\zeta \sqrt{1-\beta ^2}\right) \cos ^2(\phi )-\beta ^2+1}}, 
\end{align}
while for the ACM order we obtain 
\begin{align}
\label{eq:susc-VA-2d-ACM}
\chi_N^{\textrm{ACM}}&=\chi_N^{\textrm{CCM}}+\frac{\Lambda\beta^2 }{2\pi^2\alpha\vf}\sum_{\zeta=\pm}\int_0^{2\pi} d\phi\, 
\,  \frac{\zeta\sec^2\phi}{\sqrt{  \alpha\left(\alpha -2\zeta \sqrt{1-\beta ^2}\right) \cos ^2(\phi )-\beta ^2+1}}.
\end{align}
Since the integrand in the last term in Eq.~\eqref{eq:susc-VA-2d-ACM} (after taking the sum over $\zeta=\pm$) is positive for any 
$|\alpha|<1$ and $|\beta|<1$, with the spectrum of the corresponding Hamiltonian purely real,  we obtain that $\chi_N^{\textrm{ACM}}>\chi_N^{\textrm{CCM}}$, as also shown in Fig.~\ref{susc 2d s}. Therefore, the velocity-anisotropic tilted NH Dirac semimetal in $d=2$ shows an enhanced propensity toward the ACM ordering as compared to the CCM class, as it is the case in $d=3$, see Sec.~\ref{sec:susceptibilities}.

\section{Renormalization group analysis }
\label{a:renormalization}
In this section, we present the details of the RG analysis leading to the form of the RG flow equations of the fermionic and bosonic velocities, and the tilt parameter,  presented in Sec.~\ref{sec:NH-critical}. We employ  the leading order $\epsilon$-expansion about $d=3$ upper critical dimension of the NH Gross-Neveu-Yukawa (GNY) theory with $\epsilon=3-d$, and within the minimal subtraction scheme~\cite{ZinnJustin2002,Peskin2019}. We compute the self energy diagrams of fermions and bosons as shown in Fig.~\ref{fig:self-energy}(a) and Fig.~\ref{fig:self-energy}(b), respectively, with the form of the Yukawa interaction given in Eq.~\eqref{gny}.


\begin{figure}[t!]
    \centering
    \includegraphics[scale=0.6]{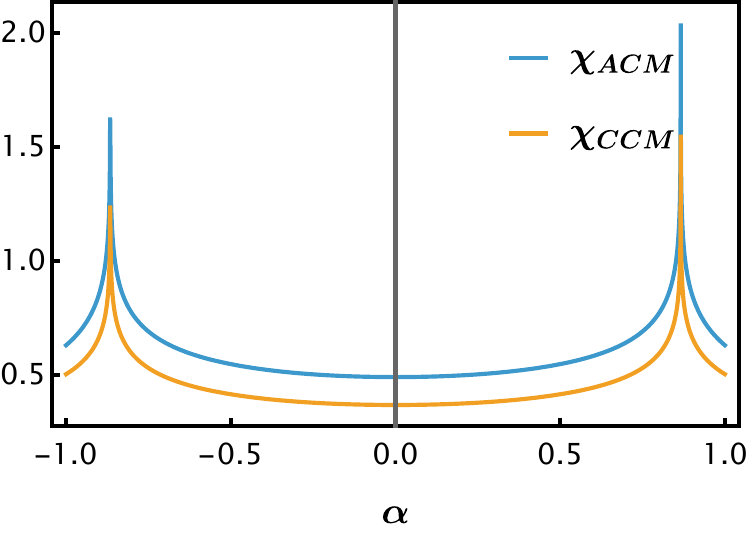}
    \caption{Mean field susceptibilities of the  noninteracting NH Dirac Hamiltonian in $d=2$ with the  velocity-anisotropy term. The susceptibility for  the commuting (anticommuting) class mass is given by Eq.~\eqref{eq:susc2dccm} (Eq.~\eqref{eq:susc2dacm}). We notice that the susceptibility has  cusps when 
    $\abs{\alpha}= \sqrt{1-\beta^2}$. The value of the anisotropy parameter $\alpha$ corresponding to the cusps coincide with the maximum slope of the density of states, as given by  Eq.~\eqref{eq:DOS-symmetric},  with the effective velocity in 
 Eq.~\eqref{eq:spectrum-velocity-anisotropy}. The values of the parameters are the same as in Fig.~\ref{susc 2d as}. }
    \label{susc 2d s}
\end{figure}


We here show the technical details for the antisymmetric tilt case, yielding the RG flow equations~\eqref{eq:beta-function-vh}--\eqref{eq:beta-function-vb} (see also Fig.~\ref{fig: assym}). The results for the velocity-anisotropic case   are obtained using the same procedure, and with the plot of  the RG flows of the velocity parameters displayed in Fig.~\ref{fig: symm}. However, we omit the details of the  calculation for the sake of brevity. 
\subsection{Fermionic self-energy}
We first calculate the fermion self-energy, with the corresponding diagram shown in Fig.~\ref{fig:self-energy}(a),
\begin{align}
    \Sigma(i \nu ,\bfk )= g^2 \int [d\omega] \int[d^d\bfq]\,\qty[N_i G_F(i\omega,q)N_i G_B(i\nu-i\omega,\bfk-\bfq)],
\end{align}
with $\int[d\omega]\equiv \int_{-\infty}^\infty\frac{d\omega}{2\pi}$ and $\int[d^d\bfq]\equiv\int \frac{d^d \bfq}{(2\pi)^d}$.
We then  obtain 
\begin{align}
     &\Sigma(i \nu , \bfk )= g^2N_i  \int [d\omega] \int[d^d\bfq]\,\Big[ -(i \omega+(\vh +\vnh M)(\bm{\Gamma}\cdot {\bf q})+\alpha \vh \Tas q_x)A_+(i\omega,\bfq) \nonumber\\
    &+ B(i\omega,\bfq)(\vh+\vnh M)(\bm{\Gamma}\cdot \bfq)\Tas q_x+2\alpha \vh q_x(\vf^2 q^2 \Tas +\alpha \vf q_x(\vh+\vnh M)(\bm{\Gamma}\cdot \bfq) )\Big] \nonumber\\ & \times N_i \frac{1}{\qty[(A_-(i\omega,\bfq))^2 -(B(i\omega,\bfq))^2)((\nu-\omega)^2+\vB^2(\vb{k}-\vb{q})^2]},
\end{align}
where $A_\pm(i\omega,\bfq)$ is given by Eq.~\eqref{eq:Aplusminus}.
  For both classes of the mass orders, we find
\begin{align}
\label{eq-a:fermion-SE}
     &\Sigma(i\nu, {\bf k})= g^2n  \int [d\omega] \int[d^d\bfq]\,\Big[ -i \omega A_+(i\omega,\bfq) +(\vh \pm\vnh M)(\bm{\Gamma}\cdot \bfq)(A_+(i\omega,\bfq)\nonumber \\
    &-B(i\omega,\bfq)\Tas  -2\alpha^2 \vh^2 q_x^2)+\alpha\vf\Tas q_x(-A_+(i\omega,\bfq)+2\vf^2 q^2)\Big]\nonumber\\
    \times  &\frac{1}{\qty[(A_-(i\omega,\bfq))^2 -(B(i\omega,\bfq))^2]\qty[(\nu-\omega)^2+\vB^2(\vb{k}-\vb{q})^2]},
\end{align}
where  the upper (lower) sign corresponds to CCM (ACM) order. 

\subsubsection{Fermionic field renormalization}


To calculate the fermion field renormalization, we extract the term proportional to the identity matrix in the self-energy, and we set external momentum ${\bf k}=0$,
\begin{align}
    \Sigma_{00}(i \nu ,{\bf k}=0 )=-g^2 n  \int [d\omega] \int[d^d\bfq]\, \frac{i \omega A_+(i\omega,\bfq)}{\qty[A_-(i\omega,\bfq))^2 -(B(i\omega,\bfq))^2]\qty[(\nu-\omega)^2+\vB^2q^2]}.
\end{align}
The renormalization condition reads
\begin{align}
 Z_\psi=1+   \pdv{ \Sigma_{00}(i \nu ,k )}{(i \nu )}\eval_{i\nu=0,\bfk=0},
\end{align}
and  after integrating over the Matsubara frequency, we find 
\begin{align}
    \pdv{ \Sigma_{00}(i \nu ,\bfk )}{(i \nu )}\eval_{i\nu=0,\bfk=0}&=-g^2n \int[d^d\bfq]\, \frac{\alpha \vh^2 q_x^2+q^2 (\vB+\vf)^2}{2q \vB(\alpha^2  \vh^2 q_x^2 -q^2(\vB+\vf)^2)^2}\nonumber\\
    &=- \frac{ng^2}{(4\pi)^2} G(\alpha,\vB,\vf) \int \frac{dq}{q},
    \end{align}
    where 
    \begin{align}
    G(\alpha,\vB,\vf)&= \int_0^{2\pi}  \frac{d\phi}{\pi}\int_0^\pi d\theta \frac{\sin \theta(\alpha^2 \vh^2 \cos^2 \phi \sin^2 \theta +(\vB+\vf)^2)}{\vB(-\alpha^2 \vh^2 \cos^2 \phi \sin^2 \theta+(\vB+\vf)^2)^2}\nonumber\\&= \frac{4}{\vB((\vB+\vf)^2-\alpha^2 \vh^2)}.
\end{align}
We then invoke  the correspondence between hard cutoff and dimensional regularization, 
\begin{align}
    \int \frac{dq}{q}\to \frac{1}{\epsilon}
\end{align}
to find for the fermion field renormalization
\begin{align}
   Z_\psi&=1- \frac{ng^2}{(4\pi)^2 \epsilon} G(\alpha,\vB,\vf). 
\end{align}
\subsubsection{Renormalization of the fermionic velocities}
To obtain  renormalization factors for fermionic velocities, we need to extract the terms proportional to the matrices $\Gamma$ in the self-energy, Eq.~\eqref{eq-a:fermion-SE}, and we choose the term proportional to $\Gamma_x$ to find the renormalization factor for the velocity $\vh$,
\begin{equation}
    \Sigma_{\Gamma_x}(i \nu ,k )= ng^2  \int [d\omega] \int[d^d\bfq]\,\Bigg[\frac{ \vh  q_x (A_+(i\omega,\bfq )-2\alpha^2 \vh^2 q_x^2)}{\qty[A_-(i\omega,\bfq)^2 -B(i\omega,\bfq)^2]\qty[(\nu-\omega)^2+\vB^2(\vb{k}-\vb{q})^2]} \Bigg].
\end{equation}
We then obtain
\begin{align}
     &\pdv{\Sigma_{\Gamma_x}(i \nu ,\bfk )}{(\vh k_x)}\eval_{i\nu=0,\bfk=0}=\frac{n g^2}{(4 \pi )^2 \epsilon}J(\alpha, \vB, \vf),
     \end{align}
     with 
     \begin{align}
    &J(\alpha, \vB, \vf) =\frac{1}{\pi} \int[d\phi d\theta] \frac{\cos^2 \phi \sin^2 \theta (-\alpha^2 \vh^2 \cos^2 \phi\sin^2 \theta+ (\vB+ \vf)^2(\vf +2\vB))}{\vB \vf (-\alpha^2 \vh^2 \cos^2 \phi \sin^2 \theta+(\vB+\vf)^2)^2}\nonumber\\
    &=\Bigg[\alpha ^2 \vh \left(-\vB \vf^2 +\vB^2 \vf+\vB^3+\alpha ^2 \vf \vh^2-\vf^3\right) \nonumber\\
    &\quad -| \alpha |  \left(\alpha ^2 \vh^2 \left(\vf^2-\vB^2\right)+ \left(\vB-\vf\right) \left(\vB+\vf\right){}^3\right)  \cosh ^{-1}\left(\frac{\vB+\vf}{\sqrt{\left(\vB+\vf\right){}^2-\alpha ^2 \vh^2}}\right)\Bigg] \nonumber\\ &\quad\times \frac{4}{\alpha ^4 \vB \vf \vh^3 \left(\vB+\vf-\alpha  \vh\right) \left(\vB+\vf+\alpha  \vh\right)}.
\end{align}
The renormalization condition for the Hermitian velocity ($\vh$) reads
\begin{align}
    Z_\psi Z_{\vh }+\pdv{\Sigma_{\Gamma_x}(i \nu ,\bfk )}{(\vh k_x)}\eval_{i\nu,\bfk=0}&=1,
     \end{align}
     implying that 
     \begin{align}
     Z_\psi Z_{\vh }&=1- \frac{n g^2}{(4 \pi )^2 \epsilon}J(\alpha, \vB, \vf),
     \end{align}
     which finally yields  the $Z-$factor for the Hermitian velocity, 
     \begin{align}
     Z_{\vh}&=1+\frac{g^2 n }{(4\pi)^2\epsilon} [ G(\alpha, \vB,\vf)- J(\alpha, \vB,\vf)].
\end{align}
Notice that the factors $
    Z_\psi$ and  $Z_{\vh }$ are independent of the class of the mass orders. 
    
To find the renormalization factor for the NH velocity, $\vnh $, we need to consider the terms proportional to the product $\Gamma M$ in the self-energy [Eq.~\eqref{eq-a:fermion-SE}],
\begin{align}
       \Sigma_{\Gamma_x M}(i \nu ,\bfk )&=\pm ng^2 \vnh  \int [d\omega] \int[d^d\bfq]\, \Bigg[\frac{q_x (A_+(i\omega,\bfq )-2\alpha^2 \vh^2 q_x^2)}{\qty[A_-(i\omega,\bfq)^2 -B(i\omega,\bfq)^2]}\nonumber \\& \quad \times \frac{1}{\qty[(\nu-\omega)^2+\vB^2(\vb{k}-\vb{q})^2]}\Bigg] 
       \end{align}
yielding 
       \begin{align}
         \pdv{\Sigma_{\Gamma_x M}(i \nu , \bfk )}{(\vnh k_x)}\eval_{i\nu=0, \bfk=0}&=\pm\frac{n g^2}{(4 \pi )^2 \epsilon}J(\alpha, \vB, \vf),
         \end{align}
         and therefore the $Z-$factor for the NH velocity reads 
         \begin{align}
         Z_{\vnh}&=1+\frac{g^2 n }{(4\pi)^2\epsilon} [ G(\alpha, \vB,\vf)\mp J(\alpha, \vB,\vf)].
\end{align}
Notice that the factor $ Z_{\vnh}$ does  depend on the  class of the mass order, with the upper (lower) sign corresponding to the CCM (ACM) OP.
\subsubsection{Renormalization of the tilt parameter}
To find the tilt renormalization factor,  we extract the terms proportional to $\Tas$, and we set $\vb{k}=k_x \vu{e}_x$ in the fermion self-energy in Eq.~\eqref{eq-a:fermion-SE},
\begin{align}
       \Sigma_{\Tas}(i \nu, {\bf k})= g^2 n  \int [d\omega] \int[d^d\bfq]\, \frac{\alpha\vh q_x (-A_+(i\omega,q )+2\vf^2 q^2)}{\qty[A_-(i\omega,{\bf q}))^2 -(B(i\omega,{\bf q}))^2]\qty[(\nu-\omega)^2+\vB^2(\vb{k}-\vb{q})^2]}.
\end{align}
After integrating over Matsubara frequencies, we obtain
\begin{align}
    &\pdv{\Sigma_{\Tas}(i \nu ,{\bf k} )}{(\vf k_x)}\eval_{i\nu=0,\bfk=0}=\frac{g^2 n \alpha}{(4\pi)^2\epsilon} I(\alpha, \vB,\vf),
    \end{align}
where, after the angular integrations, we obtain
    \begin{align}
    &I(\alpha, \vB,\vf)
    &=\frac{4  }{\alpha^2  \vB \vh^3}\left(\frac{\alpha ^2 \vh^3-\vf \vh \left(\vB+\vf\right)}{ \left(\vB+\vf\right){}^2-\alpha ^2 \vh^2}+\frac{\vf}{|\alpha| } \cosh ^{-1}\left(\frac{\vB+\vf}{\sqrt{\left(\vB+\vf\right){}^2-\alpha ^2 \vh^2}}\right)\right).
\end{align}
The renormalization condition for the tilt parameter reads 
\begin{align}
    Z_\psi Z_\alpha  Z_{\vh}\alpha \vh k_x + \vh k_x \pdv{\Sigma_{\Tas}(i \nu ,{\bf k} )}{(\vh k_x)}\eval_{i\nu,{\bf k}=0}&=\alpha \vh k_x,
    \end{align}
    and we find
  \begin{align}  
    Z_\psi Z_\alpha Z_{\vh}&=1-\frac{1}{\alpha} \pdv{\Sigma_{\Tas}(i \nu ,{\bf k})}{(\vh k_x)}\eval_{i\nu=0,k=0}= 1- \frac{g^2 n }{(4\pi)^2\epsilon} I(\alpha, \vB,\vf),
    \end{align}
    which implies the form of the renormalization parameter for the tilt 
    \begin{align}
    Z_\alpha &=1+\frac{g^2 n }{(4\pi)^2\epsilon} [ J(\alpha, \vB,\vf)- I(\alpha, \vB,\vf)]. 
\end{align}
\subsection{Bosonic self-energy}
We here calculate the bosonic self-energy, with the  Feynman diagram shown in Fig.~\ref{fig:self-energy}(b). To facilitate the calculation, we furthermore take $\vb{k}=k_x\vu{e}_x$, so that the self-energy reads
\begin{equation}
    \Pi(i\nu, {\bf k})=-\frac{g^2}{2}  \int [d\omega] \int[d^d\bfq]\,\Tr\qty[N_i G_F(i \omega,{\bf q})N_i G_F(i\nu +i \omega,{\bf k}+{\bf q})].
    \label{eqApp:Bosonic-SE}
\end{equation}
Using this expression for the bosonic self-energy, we subsequently calculate the bosonic field renormalization and the bosonic velocity renormalization factors for both the CCM and ACM OPs. 
\subsubsection{Bosonic field renormalization}
 The boson field renormalization is calculated using:
\begin{align}
    \pdv{\Pi(i\nu,{\bf k})}{\nu^2}\eval_{i\nu,{\bf k}=0}=&-\frac{g^2 n}{4\pi^ 2\epsilon} F(\alpha,\vB,\vf).
\end{align}
For the CCM order, we find
\begin{align}
  &  F_{\textrm{CCM}}(\alpha,\vB,\vf)= \frac{1}{4} \int[d\phi d\theta]\frac{\csc^2 (\theta ) \sec ^2(\phi )}{16 } \Bigg(\frac{4 \sin ^2(\theta ) \cos ^2(\phi )+2}{\vf^3}\\&+\left(\frac{2 \sin ^2(\theta ) \cos ^2(\phi )-1}{\left(\vf-\alpha  \sin (\theta ) \vh \cos (\phi )\right){}^3}+\frac{2 \sin ^2(\theta ) \cos ^2(\phi )-1}{\left(\vf+\alpha  \sin (\theta ) \vh \cos (\phi )\right){}^3}\right)\Bigg),\nonumber
    \end{align}
    which further simplifies to
    \begin{align}
     F_{\textrm{CCM}}(\alpha,\vB,\vf)=& \frac{1}{16 \vf^3 \left(\frac{\alpha^2 \vh^2}{\vf^2}-1\right)^2} \left(\frac{3 |\alpha| \vh \left(\vf^2-\alpha^2 \vh^2\right)^2}{\vf^5} \ln \left(\frac{2 \vf}{|\alpha| \vh+\vf}-1\right)\right. \nonumber\\
     &\left.-\frac{\vh^4}{\vf^4}-\frac{14 \alpha^2 \vh^2}{\vf^2}+8\right).
\end{align}
For the ACM case, following the analogous steps, we obtain 
\begin{align}
   &  F_{\textrm{ACM}}(\alpha,\vB,\vf)=\frac{ \vh }{8 \vf^5 \left(\frac{\alpha ^2 \vh^2}{\vf^2}-1\right){}^2}  \Bigg(\alpha  \vf \left(3 \left(\frac{\alpha ^2 \vh^2}{\vf^2}-1\right){}^2 \tanh ^{-1}\left(\frac{\alpha  \vh}{\vf}\right)-\frac{\alpha  \vh}{\vf}\right)\nonumber\\
     &+2 \vh \left(\frac{2 \alpha ^4 \vh^4}{\vf^4}-\frac{3 \alpha ^2 \vh^2}{\vf^2}-\frac{3 \alpha  \vh \left(\vf^2-\alpha ^2 \vh^2\right){}^2 }{\vf^5}\tanh ^{-1}\left(\frac{\alpha  \vh}{\vf}\right)+2\right)\Bigg).
\end{align}
The  renormalization condition has the form 
\begin{align}
    &Z_\phi-\pdv{\Pi(i\nu,{\bf k})}{\nu^2}\eval_{i\nu,{\bf k}=0}=1,
    \end{align}
    which yields
    \begin{align}
&Z_\phi=1-\frac{g^2 n }{4\pi^ 2\epsilon} F(\alpha,\vB,\vf).
\end{align}
Notice that the function $F(\alpha,\vB,\vf)$, and therefore also the renormalization factor $Z_\phi$ depends on the class of the mass order.  We, however, do not include this dependence in the last equation explicitly to simplify the notation. 
\subsubsection{Bosonic velocity renormalization}
To find the renormalization factor for the bosonic velocity, we need the term of the order  $k^2$ in the bosonic self-energy
\begin{align}
   \pdv{\Pi(i\nu,{\bf k})}{(v_B^2k^2)}\eval_{i\nu=0,{\bf k}=0}\equiv  \Pi_{\vB^2k^2}(i\nu,\bfk)\eval_{i\nu,\bfk=0}=&-\frac{ng^2 N_F}{4\pi^2 \epsilon} C(\alpha,\vB,\vf),
     \end{align}
     with the function depending explicitly on the class of the mass order, as we show in the following. 
     
     We first consider the  CCM case, which yields  
     \begin{align}
   & C_{\textrm{CCM}}(\alpha,\vB,\vf) =-\frac{1}{4} \int [d\phi d\theta] \frac{1}{8 \vB^2 \left(\vf^3 - \alpha^2 \vf \sin^2(\theta) \vh^2 \cos^2(\phi)\right)^3} \Bigg( \vf^8 \left(8 - 12 \sin^2(\theta) \cos^2(\phi)\right) \nonumber\\    & + \alpha^2 \vf^6 \vh^2 \left(14 \sin^4(\theta) \cos^4(\phi) + 3 \sin^2(\theta) \cos^2(\phi) - 9\right) \nonumber\\
    & + \alpha^4 \vf^4 \vh^4 \left(-14 \sin^6(\theta) \cos^6(\phi) + 9 \sin^4(\theta) \cos^4(\phi) + 2 \sin^2(\theta) \cos^2(\phi) + 2\right) \nonumber\\
    & + \alpha^6 \vf^2 \sin^2(\theta) \vh^6 \cos^2(\phi) \left(4 \sin^6(\theta) \cos^6(\phi) - 11 \sin^2(\theta) \cos^2(\phi) + 1\right) \nonumber\\
    & + \alpha^8 \sin^4(\theta) \vh^8 \cos^4(\phi) \left(2 \sin^2(\theta) \cos^2(\phi) + 1\right) \Bigg),
\end{align}
     and after the angular integrations, we find  
     \begin{align}
    C_{\textrm{CCM}} =& -\frac{1}{48 | \alpha |^3 \vB^2 \vf^4 \vh^3} \Bigg( 4 | \alpha | \vf \vh \left(2 \alpha^2 \vf^2 \vh^2 - 3 \vf^4 + 3 \alpha^4 \vh^4\right) \nonumber\\
    & + 3 \left(-4 \alpha^4 \vf^2 \vh^4 + 6 \alpha^2 \vf^4 \vh^2 - 2 \vf^6 + \alpha^6 \vh^6\right) \ln \left(\frac{2 \vf}{| \alpha | \vh + \vf} - 1\right) \Bigg).
\end{align}

     Similarly, for the ACM case, we obtain  
    \begin{align}
   & C_{\textrm{ACM}} = \frac{1}{24 \alpha ^3 \vB^2 \vf^6 \vh^3} \Bigg(6 \vf^8 \tanh ^{-1}\left(\frac{\alpha  \vh}{\vf}\right)-6 \alpha  \vf^7 \vh-6 \left(3 \alpha ^2+2\right) \vf^6 \vh^2 \tanh ^{-1}\left(\frac{\alpha  \vh}{\vf}\right)\nonumber\\&+4 \alpha  \left(2 \alpha ^2+3\right) \vf^5 \vh^3+12 \alpha ^2 \left(\alpha ^2+3\right) \vf^4 \vh^4 \tanh ^{-1}\left(\frac{\alpha  \vh}{\vf}\right) -6 \alpha ^3 \left(\alpha ^2+2\right) \vf^3 \vh^5\nonumber\\& 6 \alpha ^6 \vh^8 \tanh ^{-1}\left(\frac{\alpha  \vh}{\vf}\right)-3 \alpha ^4 \left(\alpha ^2+8\right) \vf^2 \vh^6 \tanh ^{-1}\left(\frac{\alpha  \vh}{\vf}\right)\Bigg). \label{eq: cacm}
\end{align}
The renormalization condition for the bosonic velocity reads 
\begin{align}
    Z_\phi Z_{\vB^2}- \Pi_{\vB^2k^2}(i\nu,\bfk)\eval_{i\nu,\bfk=0}&=1, \end{align}
     which yields
     \begin{align}
     Z_\phi Z_{\vB^2}&=1-\frac{ng^2 N_F}{4\pi^2 \epsilon} C(\alpha,\vB,\vf).
     \end{align}
    Finally, we find the renormalization factor for the bosonic velocity 
     \begin{align}
     Z_{\vB^2}&=1+\frac{ng^2 N_F}{4\pi^2 \epsilon} [F(\alpha,\vB,\vf)-C(\alpha,\vB,\vf)].
\end{align}
Again, both functions $F(\alpha,\vB,\vf)$ and $C(\alpha,\vB,\vf)$ depend on the class of the mass order, implying the same feature for the renormalization factor $Z_{\vB^2}$,   but we do not include this dependence here explicitly to simplify the notation. 
\end{widetext}
~
\clearpage
~
\subsubsection{ Anisotropy of the bosonic velocities}
 \begin{figure*}[t!]
        \centering
    \includegraphics[scale=0.6]{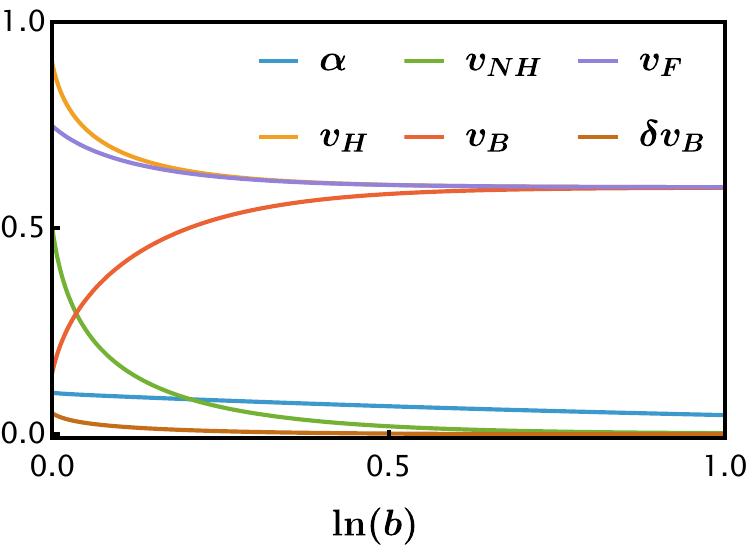}
        \caption{Renormalization-group (RG) flow of the velocity anisotropy ($\delta_{\vB}$), given by Eq.~\eqref{eqApp:bosonic-v-anisotropy} in the critical plane ($m_{ B}=0$) of the Gross-Neveu-Yukawa  theory for the quantum-critical tilted non-Hermitian  Dirac fermions close to the quantum-critical point separating the Dirac semimetal and the phase where  anticommuting class mass order parameter condenses. The RG $\beta-$functions in the theory are given by Eqs.~\eqref{eq:beta-function-vh}-\eqref{eq:beta-function-vb}.  We set the bare values of the velocity parameters $\, \vh^0=0.9,\, \vnh^0=0.5,\, \vB^0=0.15,\,\delta_{\vB}^{(0)}=0.05 $, the Yukawa coupling $g=1$, and the number of the order-parameter components $n=1$. The bare values of the tilt parameter are shown in the inset. }
      \label{fig: deltav}
    \end{figure*}  

We now show that an anisotropy in the bosonic velocities, generated by the tilt term, is in fact an irrelevant operator close to the  QCP. We here only consider the ACM order parameter,  while the CCM case can be treated analogously.
The form of such a velocity anisotropy in the renormalized theory~\eqref{eq:SR-full} is
\begin{equation}
   S_{\delta v_{B,x}^2}=-Z_{\Phi}Z_{\vB^2}Z_{\delta_{v_{B,x}^2}}\delta_{v_{B,x}^2}\vB^2\int d\tau d^d{\vb{r}} \qty(\sum_{j=1}^n\Phi_j(\partial_x^2)\Phi_j )
\end{equation}
where $\delta_{v_{B,x}^2}$ is the velocity-anisotropy parameter, and 
$Z_{\delta_{v_{B,x}^2}}$ is the corresponding renormalization factor. The renormalization condition for the anisotropy parameter reads
\begin{equation}
  Z_\phi Z_{\vB^2}Z_{\delta_{v_{B,x}^2}}- \Pi_{\vB^2k^2}(i\nu,\bfk)\eval_{i\nu,\bfk=0}=1,
\end{equation}
with the polarization bubble given by Eq.~\eqref{eqApp:Bosonic-SE}.

From this renormalization condition, we find the $\beta$-function for the anisotropy of the bosonic velocities, $\delta_{v_{\rm B}}$
\begin{widetext}
    
\begin{equation}
    \beta_{\delta_{v_{\rm B}}}=N_f n g^2(-\delta_{\vB}  \qty[F_{ACM}(\alpha, \vh,\vnh, \vB)-C_x(\alpha, \vh,\vnh, \vB )]+ \vB\qty[C_x(\alpha, \vh,\vnh,\vB +\delta_{\vB})-C_y(\alpha, \vh,\vnh, \vB )]),
    \label{eqApp:bosonic-v-anisotropy}
\end{equation}
where the function  $C_x\equiv C_{ACM}$ in Eq.~\eqref{eq: cacm}, and the function $C_y$ reads
    \begin{align}
   & C_{y}(\alpha,\vB,\vf) =-\frac{1}{4} \int [d\phi d\theta] \frac{\sin(\theta)}{16 q \left(\alpha ^2 v_F \sin ^2(\theta ) \cos ^2(\phi )-v_F^3\right){}^3} \Bigg( 8 v_F^8 \left(\sin ^2(\theta ) \sin ^2(\phi )-1\right) \nonumber\\    & - 4 \alpha ^6 \sin ^4(\theta ) v_H^2 \cos ^4(\phi ) \left(2 \sin ^2(\theta ) \sin ^2(\phi )+4 \sin ^2(\theta ) \cos ^2(\phi )-1\right)\nonumber\\
    & + v_F^6 \left(\alpha ^2 \left(-2 \sin ^2(\theta ) \sin ^2(\phi )+5 \sin ^4(\theta ) \sin ^2(2 \phi )+12 \sin ^2(\theta ) \cos ^2(\phi )+2\right)-8 v_H^2 \left(4 \sin ^2(\theta ) \sin ^2(\phi )-3\right)\right) \nonumber\\
    & + \alpha ^4 v_F^2 \sin ^4(\theta ) (2 \alpha ^2 \cos ^4(\phi ) \left(2 \sin ^2(\theta ) \sin ^2(\phi )+\sin ^4(\theta ) \sin ^2(2 \phi )+2 \sin ^2(\theta ) \cos ^2(\phi )-1\right)+v_H^2 (5 \sin ^2(2 \phi )\nonumber\\
    &-8 \cos ^4(\phi ) \left(\sin ^2(\theta ) \sin ^2(\phi )-5\right))) - \frac{1}{2} v_F^4 (\alpha ^4 \sin ^4(\theta ) \left(8 \cos ^4(\phi ) \left(5 \sin ^2(\theta ) \sin ^2(\phi )+2\right)+5 \sin ^2(2 \phi )\right)\nonumber\\
    & +\alpha ^2 v_H^2 \left(-8 \sin ^2(\theta ) \sin ^2(\phi )-4 \sin ^4(\theta ) \sin ^2(2 \phi )+96 \sin ^2(\theta ) \cos ^2(\phi )+8\right)) \Bigg).
\end{align}

After carrying out  the  angular integrations, we eventually find

   \begin{align}
    C_{y} =& -\frac{\alpha^5}{48   \alpha ^7 v_B^2 v_F^4 v_H^3 \left(v_F^2-\alpha ^2 v_H^2\right)} \Bigg( \frac{6 v_F^6 \left(v_F^2-2 v_H^2\right) \cosh ^{-1}\left(\frac{v_F}{\sqrt{v_F^2-\alpha ^2 v_H^2}}\right)}{| \alpha | }-\alpha ^2 \left(17 \alpha ^2+60\right) v_F^3 v_H^5 \nonumber\\
    & -3 | \alpha |  v_H^2 \left(2 v_H^2-v_F^2\right) \left(\alpha ^2 v_F^2 v_H^2-4 v_F^4+\alpha ^4 v_H^4\right) \cosh ^{-1}\left(\frac{v_F}{\sqrt{v_F^2-\alpha ^2 v_H^2}}\right)\nonumber\\
    & +2 v_F v_H \left(\left(13 \alpha ^2+6\right) v_F^4 v_H^2-3 v_F^6+21 \alpha ^4 v_H^6\right)\Bigg).
\end{align}
\end{widetext}
The resulting flow of the velocity anisotropy is  displayed in Fig.~\ref{fig: deltav}, which is obtained from the $\beta-$function of the   anisotropy parameter, Eq.~\eqref{eqApp:bosonic-v-anisotropy}, together with  the $\beta-$funcitons of  other velocities and the tilt parameter, Eqs.~(\ref{eq:beta-function-vh})-(\ref{eq:beta-function-vb}), which   clearly shows that this parameter is an irrelevant perturbation close to the QCP.

\bibliography{PRB-paper}
\end{document}